\documentclass[12pt,preprint]{aastex}

\input epsf.tex

\def\kmsec{\mbox{km~s$^{\rm -1}$}}
\def\teff{\mbox{T$_{\rm eff}$}}
\def\vt{\mbox{v$_{\rm t}$}}
\def\BmV0{\mbox{(B-V)$^{\rm o}$}}
\def\VmK0{\mbox{(V-K)$^{\rm o}$}}
\def\MV0{\mbox{M$_{\rm V}^{\rm o}$}}

\def\eg{\mbox{{\it e.g.}}}

\def\third{{3$^{\rm rd}$}}
\def\deg{{$^{\circ}$}}
\def\bd17{\mbox{BD +17\deg 3248}}

\begin{document}

\title{ 
The Chemical Composition  and Age of the Metal-Poor Halo Star
 \bd17\footnote{Based on observations made at three facilities: 
(a) the NASA/ESA {\it Hubble Space Telescope}, obtained at the
Space Telescope Science Institute (STScI), which is operated by the 
Association of Universities for Research in Astronomy, Inc., under NASA 
contract NAS 5-26555; (b) the Keck~I Telescope of the W. M. Keck Observatory, 
which is operated by the California Association for Research In Astronomy 
(CARA, Inc.) on behalf of the University of California and the California 
Institute of Technology; and (c) the H. J. Smith Telescope of McDonald 
Observatory, which is operated by The University of Texas at Austin.}}


\author{
John J. Cowan\altaffilmark{2},
Christopher Sneden\altaffilmark{3},
Scott Burles\altaffilmark{5},
Inese I. Ivans\altaffilmark{3},
Timothy C. Beers\altaffilmark{6}, \\
James W. Truran\altaffilmark{4},
James E. Lawler\altaffilmark{7},
Francesca Primas\altaffilmark{8},
George M. Fuller\altaffilmark{9},  \\
Bernd Pfeiffer\altaffilmark{10},
and 
Karl-Ludwig Kratz\altaffilmark{10}
}

\altaffiltext{2}{Department of Physics and Astronomy,
University of Oklahoma, Norman, OK 73019; cowan@mail.nhn.ou.edu}

\altaffiltext{3}{Department of Astronomy and McDonald Observatory, 
University of Texas, Austin, TX 78712; chris@verdi.as.utexas.edu,
iivans@astro.as.utexas.edu}

\altaffiltext{4}{Department of Astronomy and Astrophysics, Enrico Fermi 
Institute, University of Chicago, 933 E. 56th Street, Chicago, IL 60637;
truran@nova.uchicago.edu} 

\altaffiltext{5}{Department of Physics, Massachusetts Institute of Technology, 
77 Massachusetts Avenue, Room 6-113, Cambridge, MA 02139-4307; burles@mit.edu}

\altaffiltext{6}{Department of Physics and Astronomy, Michigan State
University, East Lansing, MI 48824; beers@pa.msu.edu}

\altaffiltext{7}{Department of Physics, University of Wisconsin,
Madison, WI 53706; jelawler@facstaff.wisc.edu}

\altaffiltext{8}{European Southern Observatory, Karl-Schwarzschild Strasse 2, 
D-85748 Garching bei Muenchen; fprimas@eso.org}

\altaffiltext{9}{Department of Physics, University of California at
San Diego, La Jolla, CA 92093-0319; gfuller@ucsd.edu}

\altaffiltext{10}{Institut f\"ur Kernchemie, Universit\"at Mainz, 
Fritz-Strassmann-Weg 2, D-55099 Mainz, Germany; 
pfeiffer@mail.kernchemie.uni-mainz.de,
 klkratz@mail.kernchemie.uni-mainz.de}

\begin{abstract}
We have combined new high-resolution spectra obtained with the Hubble
Space Telescope (HST) and ground-based facilities to make a comprehensive
new abundance analysis of the metal-poor, halo star \bd17.
We have detected the \third\ $r$-process peak elements osmium, platinum,
and (for the first time in a metal-poor star) gold,
elements whose abundances can only be reliably determined using HST.
Our observations illustrate a pattern seen in other similar halo stars
with the abundances of the heavier neutron-capture elements,  
including the \third\ $r$-process peak elements,  
consistent with a scaled solar system $r$-process distribution.
The abundances of the lighter neutron-capture elements, including
germanium and silver, 
fall below that same scaled solar $r$-process curve, 
a result similar to that seen in the ultra-metal-poor
star CS~22892--052. 
A single site with two regimes or sets of conditions, or
perhaps two different sites for 
the lighter and heavier neutron-capture elements, might 
explain the abundance pattern seen in this star. 
In addition we have derived a reliable abundance for the radioactive
element thorium.  
We tentatively identify \ion{U}{2} at 3859~\AA\ in the spectrum of \bd17\,
which makes this the second detection
of uranium in a very metal-poor halo star.
Our combined observations cover the widest range in proton number 
(from germanium to uranium) thus far of neutron-capture elements  
in metal-poor Galactic halo stars. 
Employing the thorium and uranium abundances in comparison  with each 
other and with several stable elements,  we determine an average 
cosmochronological  age for \bd17\
of 13.8 $\pm$ 4  Gyr, consistent with that found for  
other similar metal-poor halo stars.

\end{abstract}


\section{INTRODUCTION}

The abundance distributions in Galactic halo stars indicate the extent of
early Galactic nucleosynthesis and the nature of the earliest progenitor
stars.  
Observations of neutron-capture elements, in particular, provide
a number of important clues about the early history, chemical evolution,
and age of the Galaxy.  
There have been a number of observational and theoretical studies 
of the metal-poor halo stars, for example, that have established that 
the heavy neutron-capture ($n$-capture) elements were generally synthesized 
early in the history of the Galaxy by the rapid($r$)-process 
(Spite \& Spite 1978; Truran 1981; Sneden \& Parthasarathy 1983;
Sneden \& Pilachowski 1985; Gilroy et~al. 1988; Gratton \& Sneden 1994;
McWilliam et~al. 1995a, b; Cowan et~al. 1995; Sneden et~al. 1996, hereafter 
SMPCBA; Ryan, Norris \& Beers 1996).\nocite{spi78,tru81,sne83,sne85,
gil88,gra94,MPSS95b,cow95,sne96,ryan}
Evidence for substantial amounts of slow($s$)-process $n$-capture products 
are confined mainly to very carbon-rich stars (see Aoki et~al. 2001 and
references therein).

Additional abundance studies of the halo stars have been employed to 
study the chemical evolution of the Galaxy. 
Of particular interest is the scatter in the mean level
of the $n$-capture element to iron abundance ratios
in stars of very low metallicity 
([Fe/H]\footnote{
We adopt the usual spectroscopic notations that
[A/B]~$\equiv$ log$_{\rm 10}$(N$_{\rm A}$/N$_{\rm B}$)$_{\rm star}$~--~log$_{\rm 10}$(N$_{\rm A}$/N$_{\rm B}$)$_{\odot}$, and
that log~$\epsilon$(A)~$\equiv$ log$_{\rm 10}$(N$_{\rm A}$/N$_{\rm H}$)~+~12.0,
for elements A and B. 
Also, metallicity will be assumed here to be equivalent to the stellar 
[Fe/H] value.} $<$ --2.0). 
First noted by Gilroy et~al. (1988), 
and  confirmed by McWilliam et~al. (1995a, b) and Burris et~al. (2000;
hereafter BPASCR), 
these observations suggest 
an early chemically inhomogeneous Galaxy. In contrast, surveys of Galactic 
disk stars with [Fe/H] $>$ --1 (Edvardsson et~al. 1993; Woolf, 
Tomkin, \& Lambert  1995) 
show little scatter in $n$-capture bulk abundance levels.  
Other chemical evolution studies have attempted to understand the changing
nature of element production as a function of metallicity and time throughout
the Galaxy's history (see Wheeler, Sneden, \& Truran 1989, and references 
therein). 
Thus, while barium appears to be synthesized by the 
$r$-process at earliest times, this element is predominantly produced in solar
system material by the $s$-process (see BPASCR) in low-to-intermediate 
mass stars (Arlandini et~al. 1999\nocite{arl99}). 
Products of the $s$-process first appear in ordinary halo stars
somewhere below a metallicity of --2, with the onset of
the bulk of Galactic $s$-processing occurring near [Fe/H] = --2,
presumably at an age equivalent to the stellar evolutionary lifetime
of low-to-intermediate mass stars (BPASCR).

The abundances of certain long-lived radioactive elements, 
such as thorium and uranium, act as nuclear chronometers 
and provide direct age determinations in ultra-metal-poor halo stars 
(those with [Fe/H]~$\lesssim$ --2.5, hereafter UMP stars).
This technique has been used recently with Th (produced solely in the 
$r$-process), particularly with respect to the stable $r$-process element 
europium, in both halo field stars (SMPCBA; Cowan et~al. 1997; 
Pfeiffer, Kratz, \& Thielemann 1997; Cowan et~al. 1999; 
Sneden et~al. 2000a; Westin et~al. 
2000; Johnson \& Bolte 2001\nocite{sne96,cow97,pfe97,cow99,sne00a,wes00,jb01}) 
and globular cluster stars (Sneden et~al. 2000b\nocite{sne00b}). 
To date, both nuclear chronometers Th and U have been detected in only
one UMP star (Cayrel et~al. 2001\nocite{cay01}).
Radioactive-element-age
    estimates depend sensitively 
upon both theoretical 
nucleosynthesis predictions and accurate determinations 
of the stellar abundances of radioactive elements. 
Thus far the reliability of the technique, indicated by still large 
uncertainties in chronometric age estimates,
has been limited by very  few stellar detections of Th and/or U and 
correspondingly by few detections of the stable \third\ $r$-process peak 
elements necessary to constrain theoretical predictions.  

Despite surveys of large numbers of halo stars  
(Beers, Preston, \& Shectman 1985, 1992) 
only a few UMP stars have been selected for 
extensive followup abundance studies. 
These include the now well-known UMP giant star CS~22892-052 
(Sneden et~al. 1994, 1996, 2000a; Cowan et~al. 1995, 
1997\nocite{sne94,cow95,sne96,cow97,sne00a}). 
While a great many elements have been
detected in CS~22892-052 -- second only to the Sun --  
abundance studies have been limited to ground-based observations. 
(Results from space-based observations of CS~22892--052 will 
be given in Sneden et~al. 2001, in preparation.)
This has prevented reliable abundance measures for the important
\third\ $r$-process peak
elements Os and Pt,  which have their strongest atomic 
transitions in the UV. 
These elements, along with lead, were first detected in a metal-poor 
halo star by Cowan et~al. (1996). 
Uranium has only recently been found, for the first time, in the 
UMP star CS~31082-001 ([Fe/H]~= --2.9, Cayrel et~al. 2001).
Westin et~al. (2000) were able to obtain \third\ $r$-process peak element
abundances with the Hubble Space Telescope (HST), along with thorium, 
in the UMP ([Fe/H] = --3.0) star HD~115444. 
This was the first halo star with reliable abundance
data on the heaviest stable elements and the radioactive element thorium.
However, the other important nuclear chronometer, uranium,
has not yet been detected in that star.

In this paper we present the results of an extensive series of
observations regarding the metal-poor ([Fe/H]~= --2.1) halo star \bd17. 
A separate investigation of this star is of interest for three reasons.
First, \bd17\ has relatively large ratios of $n$-capture elements to iron 
(see e.g., BPASCR), which made additional element detections likely. 
These very large relative $n$-capture overabundances in \bd17\ are similar 
to those in HD 115444, while being somewhat less than those in CS~22892--052
and CS~31082-001.
Second, the overall [Fe/H] metallicity of \bd17\ is roughly a factor of 10 
larger than those of the other very $n$-capture-rich stars.
Third, \bd17\ is a warmer star than the others, by 400-500~K in \teff.
Thus, \bd17\ was targeted as (perhaps) the most metal-rich of the very
$n$-capture-rich halo stars, with somewhat different excitation/ionization
atmospheric conditions, that might provide new abundance measurements 
and also might be suitable for a chronometric age estimate.
In addition to the ground-based observations, we have obtained HST observations 
allowing a comprehensive analysis of the $n$-capture elements with the widest 
range in proton number so far in any halo star, from Ge to U. 
Our observations are presented in \S 2, followed by a summary of
the derivation of abundances in \S 3.
The significance of the derived $n$-capture abundances and application
of radioactive chronometer elements to stellar age estimates are discussed 
in \S 4, and conclusions given in \S 5.

\section{OBSERVATIONS AND REDUCTIONS}

{\em UV data:}
High resolution spectra in the ultraviolet wavelength range
2410~$\leq$ $\lambda$~$\leq$ 3070~\AA\ were obtained with the Hubble 
Space Telescope Imaging Spectrograph ($STIS$).  
The instrument was configured with echelle grating E230M centered
at $\lambda$~= 2707~\AA, an entrance aperture of 
0.2$\arcsec\times$0.06$\arcsec$, and the $NUV-MAMA$ detector, which 
combined to deliver a spectral resolving power of 
R~$\equiv$~$\lambda/\Delta\lambda$ $\simeq$~30,000.
Four individual spectra were gathered during the single HST visit for 
\bd17, with a total integration time of 180~min.
Standard HST pipeline reductions were employed to reduce the individual
integrations to one-dimensional, flat-fielded, wavelength-calibrated spectra.
We then operated on these spectra and combined them by employing the HST 
software tools of $IRAF.$\footnote{
IRAF is distributed by the National Optical Astronomy Observatories, which 
are operated by the Association of Universities for Research in Astronomy, 
Inc., under cooperative agreement with the National Science Foundation.}
The final signal-to-noise (S/N) of the spectrum was difficult to
estimate because stellar absorption features overlap at these
wavelengths; true spectral continuum windows are difficult to identify.
From comparisons of synthetic and observed spectra we believe
that S/N~$>$ 50 in the wavelength regions of interest for this study.

{\em Near-UV data:}
We employed the Keck~I $HIRES$ cross-dispersed echelle spectrograph
(Vogt et~al. 1994\nocite{Vetal94}) to acquire high-resolution spectra in 
the wavelength interval 3150~$\leq$ $\lambda$~$\leq$ 4600~\AA.
The spectral coverage was complete throughout this range.
The resolving power of the fully reduced spectrum was R~$\simeq$ 45,000.
Estimation of the S/N was not easy at the shortest wavelengths, but 
certainly the S/N decreased blueward, from $\sim$300 at 
$\lambda$~$\sim$ 4500~\AA\ to $\sim$30 at $\lambda$~$\sim$ 3200~\AA, due to 
the combined effects of decreasing stellar flux and CCD response in the UV. 
We also obtained standard auxiliary spectra of a tungsten lamp
for flat-fielding the spectrum and a Th-Ar lamp for wavelength calibration.
We performed a full echelle extraction, including flat-field and bias
correction, cosmic-ray rejection, sky-subtraction, 
wavelength and flux calibration, utilizing 
the software package MAKEE (e.g., Barlow \& Sargent 1997).

{\em Longer wavelength data:} 
We obtained a high-resolution spectrum spanning from the violet to the
near-IR with the McDonald 
Observatory 2.7m H.~J. Smith telescope and the ``2d-coud\'e'' 
cross-dispersed echelle spectrograph (Tull  et~al. 1995)\nocite{TMSL95}.
In addition to tungsten and Th-Ar calibration spectra, we also
acquired a spectrum of the hot, rapidly rotating star $\zeta$~Aql
to assist in cancelation of telluric (O$_2$ and H$_2$O) absorption
features in the yellow-red wavelength region.
With the spectrograph configured with a 1.2$\arcsec$ entrance slit and
a Tektronix 2048$\times$2048 CCD detector, the 2-pixel spectral
resolving power was R~$\simeq$ 60,000.
The useful wavelength range was 3750~$\leq$ $\lambda$~$\leq$ 9500~\AA.
The spectrum was continuous for $\lambda$~$<$ 5900~\AA, but the increasing 
free spectral range of orders longward of this wavelength created some gaps
in spectral coverage.
The S/N value of the \bd17\ spectrum after completion of the reduction 
procedures was $>$250 for $\lambda$~$>$ 5500~\AA, and declined
steadily to levels of $\sim$130 at $\lambda$~$\sim$ 4200~\AA.
Initial processing of the raw spectra (bias subtraction and frame trimming,
scattered light removal, cosmic ray excision, flat-fielding, spectral order 
extraction, and wavelength calibration) was accomplished with standard 
$IRAF$ echelle reduction tasks.  

{\em Processing the spectra:}
The final reduction tasks discussed here were accomplished with the 
specialized spectrum manipulation software package $SPECTRE$ (Fitzpatrick 
\& Sneden 1987)\nocite{FS87}.  
For all stellar data we eliminated individual bad data 
points,  and then normalized the spectral continua by fitting third-order 
spline functions to sets of interactively chosen continuum points. 
Additionally, all of the STIS spectral orders were smoothed via convolution 
with a 2-pixel FWHM Gaussian function.
Equivalent widths (EWs) of selected lines were measured either by
fitting Gaussian profiles to the spectral lines or by direct
(Simpson's Rule) integrations over the line profiles; sometimes both
methods were employed for lines that were partially blended or
had asymmetric line shapes.
The EWs of atomic species of elements with Z~$\leq$~30 were needed to
determine a new model atmosphere for \bd17\ and to confirm that this
star has a typical Pop~II abundance mix, and EWs of many species
with Z~$>$~30 were employed in the $n$-capture abundance analysis.
The line lists for these tasks are described in \S3.

\section{ABUNDANCE ANALYSIS}

\subsection{Derivation of the Model Atmosphere}

We first determined new model atmosphere parameters \teff, log~$g$, [M/H],
and \vt\ for \bd17\ using the EW data.
This was accomplished in a standard LTE analysis that sought 
to derive: $(a)$ \teff, from minimizing the line-to-line
abundance differences with excitation potential E.P. (mainly for
\ion{Fe}{1}); $(b)$ \vt, from minimizing the line-to-line abundance
differences with EW; $(c)$ log~$g$, from forcing reasonable agreement
between the neutral and ionized species of Ti and Fe; and $(d)$ [M/H],
from the approximate mean abundance level of the Fe-peak elements
and the $\alpha$-capture elements Mg, Si, and Ca.
We employed the line lists of Westin et~al. (2000)\nocite{wes00} for 
these and all other species with Z~$\leq$~30.
Trial model stellar atmospheres were generated from the 
Kurucz (1995)\footnote{see http://cfaku5.harvard.edu/}
grid with interpolation software kindly provided by A. McWilliam (private
communication). 
Abundances were computed for each line with the current version of Sneden's 
(1973)\nocite{sne73} line analysis code, iterating various atmosphere 
parameter combinations until the best overall model was achieved.

\begin{figure}
\epsscale{1.0}
\plotone{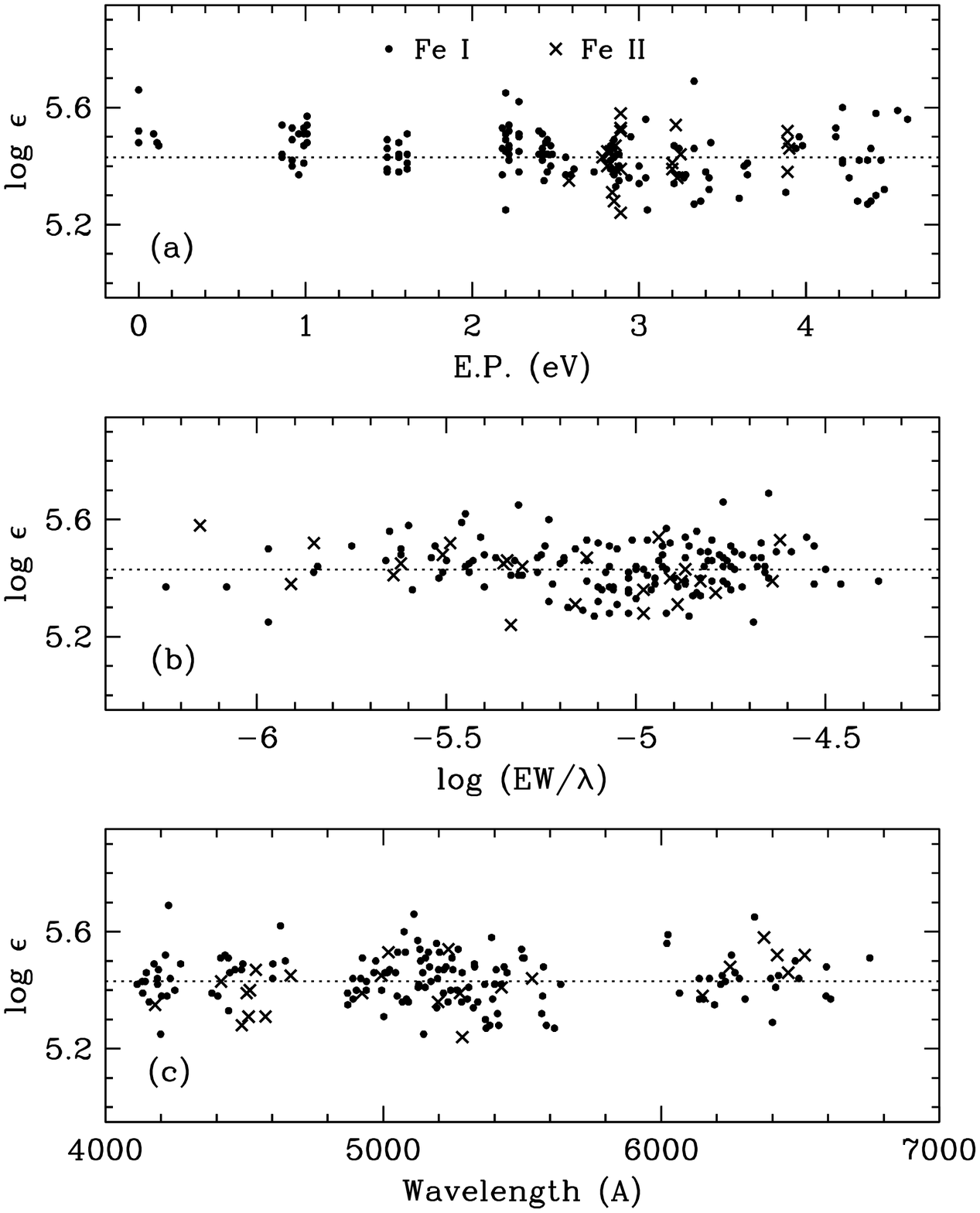}
\caption{
Abundances of \ion{Fe}{1} and \ion{Fe}{2} lines in \bd17\ as functions 
of excitation potential (panel a), logarithm of the reduced equivalent width 
(panel b), and wavelength (panel c), using the adopted model atmosphere.
\label{f1}}
\end{figure}

In Figure~\ref{f1} the correlations of \ion{Fe}{1} and \ion{Fe}{2}
lines with E.P., reduced equivalent width log(EW/$\lambda$), 
and wavelength are shown for the adopted model:
(\teff, log~$g$, [M/H], \vt)~= (5200~K, 1.80, --2.0, 1.9~\kmsec).
Within the scatter of individual line measurements ($\sigma$~= 0.08
for \ion{Fe}{1} and 0.09 for \ion{Fe}{2}) there are no obvious trends
in the Fe abundances.
Additionally, the same model atmosphere choice yields no anomalous
trends in the line-to-line abundances of any other species.
Thus, the model atmosphere is internally well determined.

For external comparisons, note first that our model atmosphere 
parameters are in good agreement with those from all but one of the
few prior spectroscopic investigations of \bd17.
Luck \& Bond (1981)\nocite{LB81} provided an initial set of
parameters from photographic coud\'e spectra:
(5000~K, 1.80, --2.0, 3.0~\kmsec).
BPASCR derived abundances for a few 
$n$-capture elements in this star, but adopted the Pilachowski, Sneden \& 
Kraft (1996) model atmosphere parameters: (5250~K, 2.30, --1.80, 1.5~\kmsec).
Those values were determined from combined photometric and (very limited) 
spectroscopic information.
Recently, Johnson \& Bolte (2001)\nocite{jb01} have obtained their own 
Keck~I HIRES spectrum of \bd17, and their model atmosphere, derived in the 
same manner as we have done, has parameters essentially identical to ours: 
(5200~K, 1.80, --1.95, 1.90~\kmsec).
The only substantial differences are seen with the investigation of
Klochkova, Ermakov, \& Panchuk (1999)\nocite{kep99}, who suggest that \bd17\ 
is a much hotter star, with corresponding increases in gravity and 
metallicity: (5590~K, 2.4, --1.15, 1.1~\kmsec).
However, these parameters seem to be incompatible with the other
high-resolution spectroscopic studies and with photometric data
for this star that will be considered next.

In comparing our model atmosphere parameters with ones derived from 
photometric indicators, caution should be employed because reddening 
estimates for \bd17\ are larger than for many halo stars ($E(B-V)$~= 0.06, 
Bond 1980\nocite{bon80}; $E(b-y)$~= 0.040, translating to $E(B-V)$~= 0.055, 
Anthony-Twarog \& Twarog 1994\nocite{att94} and private communication).
Such reddening can introduce substantial uncertainties in photometric 
temperature and gravity estimates.

Extant broad-band photometry includes UBV measures $V$~= 9.37, and 
$B-V$~= 0.66 (Carney 1983)\nocite{car83}, and IR measures $J$~= 7.815, 
$H$~= 7.400, and $K$~= 7.310 (Alonso, Arribas, \& Martinez-Roger 
1998)\nocite{aam98}.  
The $V-K$ color is often thought to be the best broad-band photometric 
temperature indicator.
Alonso, Arribas, \& Martinez-Roger (1999, 2001)\nocite{aam99} recommend
formulae for converting $V-K$ values and [Fe/H] estimates to \teff,
but there are two complications for \bd17.
First, this star has a $V-K$ in the overlap region between two
different Alonso et~al. (1999, 2001) formulae for \teff$_{V-K}$.
Second, \bd17\ has been classified as a red horizontal branch (RHB) star
(Bond 1980; Pilachowski et~al. 1996; Alonso et~al. 1998).
Thus since the Alonso et~al. formulae strictly apply to first-ascent
true red giant branch (RGB) 
stars, application of them to \bd17\ may be questionable.
Remembering these caveats, and adopting [Fe/H]~= --2.1 in the
\teff\ calculations, and (for the moment) assuming no interstellar 
reddening, then the observed $V-K$~= 2.06 yields \teff$_{V-K}$~= 4985~K or
5025~K (rounding the value to the nearest 5~K), using equations \#8 and
\#9 respectively from Table~2 of Alonso et~al. (1999, 2001). 
However, if \bd17\ is reddened according to the Bond (1980) and Anthony-Twarog 
\& Twarog (1994) estimates, and if  the intrinsic $V-K$ color may be 
estimated as $(V-K)_{int}$~$\approx$ $(V-K)_{0}-A_V$ $\approx$ 
$(V-K)-3\times E(B-V)$~= 1.88, then \teff$_{V-K}$~= 5280~K or 5260~K
from equation \#8 and \#9 respectively of Alonso et~al. 
Thus the spectroscopic and photometric temperature estimates for
\bd17\ agree well if $E(B-V)$~$\approx$ 0.06.

The JHK photometry of \bd17\ also can be employed to estimate \teff\ 
without reference to shorter wavelength photometry, thus avoiding
reddening questions.
From the Alonso et~al. (1999) equation \#10, the observed $J-H$~= 0.415 yields
\teff$_{J-H}$~= 5080~K, and from equation \#11 the observed $J-K$ yields
\teff$_{J-K}$~= 5010~K (this value is independent of assumed metallicity).
Therefore the total range in broad-band photometric temperature estimates is 
about 300~K, which happily encompasses the spectroscopic value.
The photometric \teff\ range is unlikely to be narrowed further due to the 
RHB status and the uncertain reddening of \bd17.

Photometry and a distance estimate for \bd17\ can provide an ``evolutionary''
gravity via the standard formula,
$$
\log g_{evol} = -12.50 + 0.4\left({\rm M_V} + {\rm BC}\right) +
\log \mathcal{M} + 4\log \teff\
$$
where BC is the bolometric correction and $\mathcal{M}$ is the stellar mass.
Using the spectroscopic \teff~= 5200~K, equation \#18 of Alonso et~al. (1999,
2001)
gives BC~= --0.21, and this value is consistent with other estimates (e.g.,
Bell \& Gustafsson 1978\nocite{bg78}).
The mass of \bd17\ should lie between that of a typical main 
sequence turn-off Pop~II star, 
0.85$\mathcal{M}_{\odot}$, and that of a HB star, 0.55$\mathcal{M}_{\odot}$.
An Hipparcos parallax (Perryman et~al. 1997) exists, albeit with a large 
uncertainty: $\pi$~= 3.67~$\pm$~1.50~mas.
This parallax and the $V$ magnitude correspond to
$M_V$ = +2.16$^{+0.74}_{-1.14}$
placing the star on the subgiant branch. 
This in turn implies with the above formula that 
2.4 $\lesssim$ log~$g$ $\lesssim$ 3.4, the entire range of possible values
being much larger than we determined from the \bd17\ spectrum
(see also Figure~9 of Carbon et~al. 1982\nocite{Cetal82}).
This higher gravity is at odds with the highly evolved position of 
\bd17\ in the Str\"omgren $c_1$ versus $b-y$ diagram (Figure~5 of Bond 1980),
which has been shown to be a robust indicator of evolutionary state for
cool metal-poor stars (Pilachowski, Sneden, \& Booth 1993\nocite{psb93}; 
Gratton et~al. 2000\nocite{gscb00}).
The Hipparcos parallax probably should be disregarded for this star.

Bond (1980) estimated  $M_V$~$\approx$ --0.2 for \bd17, and if this
star belongs to the RHB, then $\mathcal{M}$~= 0.55$\mathcal{M}_{\odot}$.
These values and an assumption of \teff~= 5200~K lead to 
log~$g_{evol}$~$\approx$ 1.8, in excellent agreement with the 
spectroscopic value.
If we instead assume a temperature at the high end of the estimates,
    \teff~= 5600~K (Klochkova et~al. 1999), a higher absolute magnitude
    estimate, $M_V$~$\approx$ +0.65 (Anthony-Twarog \& Twarog 1994 and
    private communication), and a main sequence mass of
    $\mathcal{M}$~= 0.85$\mathcal{M}_{\odot}$, then log~$g$~$\approx$ 2.6.
    A similar extension of parameters in the opposite direction yields
    log~$g$~$\approx$ 1.6.
This exercise, just as in the comparison of spectroscopic
and photometric temperature estimates, confirms that the spectroscopic
gravity is not far from the evolutionary value, but uncertainties
in (especially) the absolute magnitude of \bd17\ preclude more definitive
statements.

Our metallicity value can also be compared to earlier literature estimates.
From uvby photometry, Bond (1980) originally derived [Fe/H]$_{uvby}$~= --2.0,
and more recently Anthony-Twarog \& Twarog (1998)\nocite{att98} have suggested
[Fe/H]$_{uvby}$~= --2.16.
Beers \& Sommer-Larsen (1995)\nocite{BS95} estimated metallicities for a 
large sample of halo stars from a variety of photometric and low-resolution 
spectroscopic surveys in the literature, and gave [Fe/H]~= --2.0 for \bd17.
Beers et~al. (1999)\nocite{brnrs99} used broad-band photometry, 
\ion{Ca}{2} K~line EWs, and medium-resolution spectra of a large sample of 
metal-poor stars to provide a uniform [Fe/H] scale over three decades of 
stellar metallicity, and in their system they derived 
[Fe/H]$_{K-line}$~= --2.07~$\pm$~0.13 for \bd17.
Photometry on the DDO system, using the calibration of Clari\'a et~al. 
(1994)\nocite{cmpl94} and the observations of Hartkopf \& Yoss 
(1982)\nocite{HY82}, yields [Fe/H]$_{DDO}$~= --2.21~$\pm$~0.14.
The derived [Fe/H] values from the previous high-resolution spectroscopic
analyses 
(neglecting the discordant value of Klochkova et~al. 1999)
are --2.04 (Luck \& Bond 1981)\nocite{LB81}, --2.02 (Pilachowski
et~al. 1996), and --2.11 (Johnson \& Bolte 2001).
In short, the metallicity of \bd17\ appears to be well established,
as the literature photometric and spectroscopic estimates are in 
good accord with our derived [Fe/H]~= --2.09. 

Considering all the available spectroscopic and photometric information
    for \bd17, we suggest that the model atmospheric parameters for
    this star are well determined at (\teff, log~$g$, [M/H], \vt)~=
    (5200$\pm$150~K, 1.80$\pm$0.3, --2.0$\pm$0.2, 1.9$\pm$0.2~\kmsec).

\subsection{Abundances of Elements with Z~$\leq$~30}

With our final spectroscopically determined model atmosphere for \bd17\
we then computed the abundances of almost all of the non-$n$-capture
elements using the measured EWs.
Abundances of light elements Li, C, N, and O were derived from
    matching synthetic spectra to the observed features of the
    \ion{Li}{1} 6708~\AA\ resonance doublet, the CH
    $A^{\rm 2}\Delta-X^{\rm 2}\Pi$ G-band from 4358 to 4372~\AA,
    the NH $A^{\rm 3}\Pi_i-X^{\rm 3}\Sigma^-$ 0-0 and 1-1 bands from
    3355 to 3375~\AA, the CN $B^{\rm 2}\Sigma^+-X^{\rm 2}\Sigma^+$
    0-0 and 1-1 bands from 3869 to 3884~\AA, and the [\ion{O}{1}]
    6300~\AA\ line.  For details about the construction of the line
    lists employed in these computations the reader is referred to
    Sneden et~al. (1991)\nocite{skpl91}; Westin et~al. (2000);
    and Gratton et~al. (2000)\nocite{gscb00}.  The \ion{Li}{1}
    feature is extremely weak (EW~$\leq$ 1.5~m\AA\ or
    log~(EW/$\lambda$)~$\leq$ --6.6) and our detection of it in
    \bd17\ must be considered tentative.  The primary N abundance
    indicator is NH, since the CN absorption even at the 0-0 bandhead
    at 3883~\AA\ is no more than 10\% of the continuum, and the
    N abundance derived from CN depends directly on the assumed
    abundance of C.

\begin{figure}
\epsscale{1.0}
\plotone{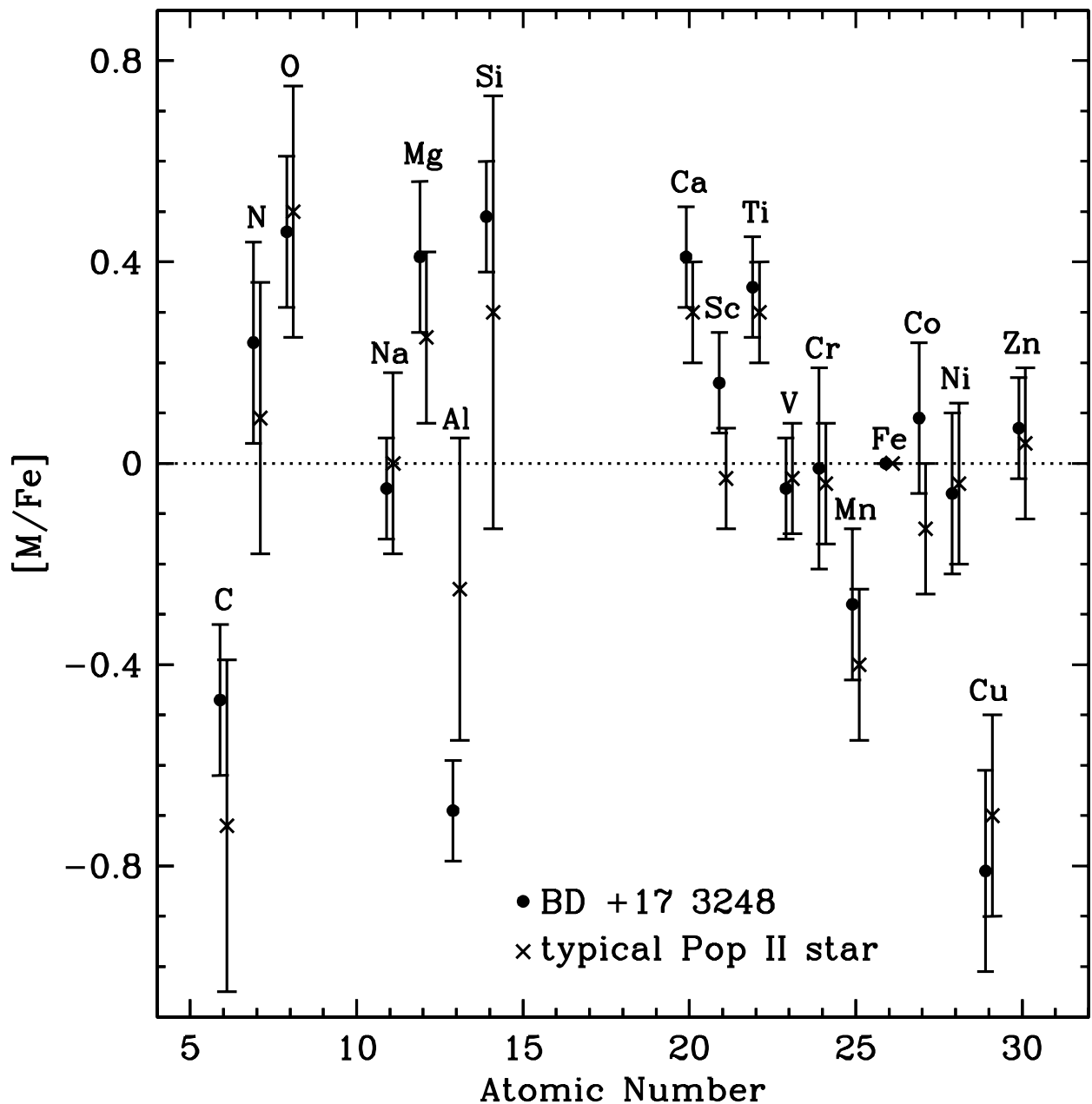}
\caption{
Abundance ratios for elements with Z~$\leq$~30 derived in this analysis
of \bd17, and mean ratios for typical Pop~II stars.
Most of the Pop~II star mean abundance ratios are taken from Figure~4
(the points for metallicity [Fe/H]~= --2.4) of Cayrel's (1996) review paper.
The mean ratios for C and N come from Kraft et~al. (1982), the value for V is
from Gratton \& Sneden (1991), and those for Cu and Zn are from
Sneden et~al. (1991).
The error bar plotted for a \bd17\ abundance is the larger of 0.10~dex or
the sample standard deviation from Table~\ref{tab1}.
\label{f2}}
\end{figure}

The derived abundances of Z~$\leq$~30 elements are given in 
Table~\ref{tab1} and graphically displayed in Figure~\ref{f2}.
Elements whose derived abundances are based on only one or two lines
(Al, V, Co, Cu, and Zn) clearly should be treated with caution.
Here we will elaborate on a few of the abundances.

First, note that the Na abundance has been derived from the
5685~\AA\ and 8190~\AA\ \ion{Na}{1} doublets.
Use of the Na~D resonance lines would yield about a factor of four larger
abundance than that determined from the higher excitation lines.
This apparent anomaly is similar to that found in RHB stars of
similar-metallicity globular clusters M92 and M15 (Sneden et~al. 
1997\nocite{setal97}; Sneden, Pilachowski, \& Kraft 2000)\nocite{spk00}.
In these clusters, the D lines and the 5680~\AA\ doublets give similar 
abundance values in true first-ascent RGB, but not RHB, stars.  
This lends inferential support to the suggested RHB status of \bd17.
The spectra of RHB stars in M92 and M15 are most easily distinguished from
those of the RGB stars by the presence of H$\alpha$ emission
(Sneden et al. 2000; Pilachowski \& Sneden 1999), but unfortunately
this feature fell in one of the gaps of our McDonald spectral order coverage.
Observations taken in 1983 as part of 
the H$\alpha$ survey of metal-poor field giants by Dupree \& Smith
(1988)\nocite{ds88} do not show H$\alpha$ core emission 
but this star was not included in their more recent survey (Dupree
\& Smith 1995\nocite{ds95}).  
R. Cayrel (private communication) finds no H$\alpha$ anomaly in their 
new VLT spectrum of \bd17.
Further comparison of chromospheric features in this star with those in 
RGB stars of comparable metallicity and absolute magnitude would be welcome.

Second, the very low Al abundance, determined from only the \ion{Al}{1} 
3961~\AA\ resonance line, may not represent the true Al content of \bd17.
It is well known that the higher excitation lines in the red spectral
region yield substantially larger Al abundances than those derived from the
resonance lines.
A good discussion of this problem and references to previous work
are given in Ryan et~al. (1996)\nocite{ryan}.
Unfortunately, none of the lines of the 6697 or 8773~\AA\ \ion{Al}{1}
doublets could be detected in our spectra of \bd17, and estimated EW
upper limits of $\approx$2~m\AA\ 
translated to an Al upper limit $\approx$1.5~dex larger than the
    abundance derived from the 3961~\AA\ resonance line.  The upper limit
    from the red lines is not helpful,
and so the Al
abundance of \bd17\ must remain ill-determined at present, although 
following the lead of Ryan et~al. we suspect that the true abundance 
may be  larger than we have determined from the
resonance line.

Finally, note that the abundance of O in metal-poor stars is
    a controversial subject at present.  Analyses of [\ion{O}{1}] and
    OH IR vibration-rotation lines yield [O/Fe]~$\simeq$+0.4 independent
    of [Fe/H] over the entire halo metallicity range (\eg, Gratton et~al.
    2000\nocite{gscb00} and references therein; Balachandran
    \& Carney 1996; Mel\'endez Barbuy \& Spite 2001\nocite{MBS01}), 
     but \ion{O}{1}
    high excitation triplet lines and OH near-UV electronic band lines
    yield linearly increasing [O/Fe] values with decreasing metallicity,
    reaching [O/Fe]~$\sim$~+1 at [Fe/H]~$\sim$~--3 (Israelian,
    Grac\'ia Lopez, \& Rebolo 1998\nocite{IGR98}; Boesgaard et~al.
    1999\nocite{BKDV99}).  In \bd17\ the [\ion{O}{1}] 6300~\AA\
    line and all 7770~\AA\ triplet lines were detected, and our analysis
    reproduced the clash between these two O abundance indicators: from
    the forbidden line we derived [O/Fe]~= +0.46 and from the triplet
    lines we found [O/Fe]~= +0.72.  In Table~\ref{tab1} we quote
    the abundance derived from the [\ion{O}{1}] line because
    that is the usual O abundance indicator for halo giant stars.
    The analysis of \bd17\ sheds no additional light on the
    general O abundance problem of metal-poor stars, and a full
    examination of this issue is beyond the scope of our study.

In order to place the Z~$<$~30 elemental abundances of \bd17\ in context
with general Pop~II abundance trends, we add to Figure~\ref{f2} 
estimates of mean [M/Fe] ratios for field stars of similar metallicity.
For most elements, these overall abundance means were taken from the 
points representing stars of [Fe/H]~$\approx$ --2.4 in Figure~4 of 
Cayrel (1996)\nocite{cay96}.
Among elements not considered in Cayrel's review, the source for the 
mean V abundance is Gratton \& Sneden (1991)\nocite{gra91},
and for Cu and Zn, Sneden, Gratton, \& Crocker (1991)\nocite{sgc91}.
The abundances of the volatile elements Li, C, and N change with
evolutionary state.  
Therefore for the Pop~II abundance trends of C and N we used mean values from
the Kraft et~al. (1982)\nocite{ketal82} survey of C and N in halo
field giant stars with metallicities more similar on average to that
of \bd17.  These means were computed considering only those Kraft
et~al. stars that are thought to be chemically evolved, by adopting 
a luminosity limit of $M_V$~$<$~--0.5.  These means are not very different 
from more recent results for evolved stars in the  higher metallicity domain
--1~$>$ [Fe/H]~$>$ --2 (Gratton et~al. 2000).  Abundances of Li are
not pictured in Figure~\ref{f2} because its origin in the halo
is due to initial Big Bang nucleosynthesis and it is destroyed in normal
stellar evolution; thus values of [Li/Fe] are not meaningful. However,
our abundance log~$\epsilon$(Li)~$\lesssim$ +0.2 is in accord with
the upper limit for field RHB stars of the Gratton et~al. survey
(log~$\epsilon$(Li)~$<$ +0.2).

The comparison of abundance sets in Figure~\ref{f2} indicates that
the abundances of elements with Z~$\leq$~30 in \bd17\ conform to the
usual Pop~II abundance pattern in this metallicity domain: elevated
abundances of $\alpha$-capture elements Mg, Si, Ca, and Ti;
large deficiencies of Al and Mn, and relatively normal abundances of
most of the Fe-peak elements.
Any deviations from this pattern are significantly less than the
uncertainties in the derived \bd17\ abundances.

\subsection{Abundances of the $n$-Capture Elements}

Abundances of 26 $n$-capture elements and a significant upper limit
    for a 27$^{th}$ one were determined from EW matches 
for unblended single lines, and from synthetic spectrum
computations for blended lines and/or lines with significant hyperfine
and/or isotopic splitting. 
The $n$-capture abundance analysis generally followed the ones described 
by SMPCBA and Sneden et~al. (2000a)\nocite{sne96,sne00a} for CS~22892-052, 
Westin et~al. (2000)\nocite{wes00} for HD~115444, and Sneden et~al. 
(1998)\nocite{sne98} for UV spectral region $n$-capture transitions in 
HD~115444 and HD~126238.
We began with the line lists that were developed for those papers.
But in the past several years significant improvements have been made to
the atomic line parameters of several $n$-capture elements, particularly
for some singly ionized species of rare-earths.
Additionally, the brief Sneden et~al. (2000a) paper on CS~22892-052 did 
not discuss transition probability choices for the new $n$-capture 
elements reported in that study.
Finally, the combination of higher temperature and metallicity of \bd17\
compared to CS~22892-052 and HD~115444 sometimes reveals different sets
of spectral features of these species for analysis than were best for
the other two stars.
Therefore we reconsidered the abundance determinations for all $n$-capture
elements, and in the Appendix we comment on those species that have had 
significant updates in line list parameters or are otherwise deserving of 
some detailed discussion.

The construction of synthetic spectrum line lists are described in 
SMPCBA and Sneden et~al. (1998).
Briefly, we culled the lists of atomic lines from the Kurucz (1995) 
database, retaining \ion{H}{1} Balmer lines and those atomic lines of 
neutral and singly ionized species whose excitation potentials were 
less than 5~eV.
Trial synthetic spectrum computations showed that only transitions 
meeting these criteria would  produce measurable absorption in our 
spectra of cool, metal-poor stars.  
We supplemented these lines with a few that are identified in the 
Moore, Minnaert, \& Houtgast (1966)\nocite{MMH66} solar atlas but 
are missing from the Kurucz database.
The relevant molecular lines are in almost all cases the hydrides CH,
NH, and OH.
Again starting from the Kurucz lists, we kept lines of these molecules
that had E.P.~$\lesssim$ 1.5, since transitions beginning at higher excitation
levels of these easily dissociated molecules are generally bound-free
transitions that proved to be undetectable in our observed spectra.
To account for a few absorptions in the spectra that have no
obvious atomic and molecular identification, we arbitrarily assumed
that they were \ion{Fe}{1} lines with E.P.~= 3.5~eV.

We refined these  line lists by matching synthetic
and observed spectra of the solar photosphere for spectral regions
covered by our ground-based Keck and McDonald spectra. 
The solar spectrum was an electronic copy of the Kurucz et~al. 
(1984)\nocite{KFBT84} solar flux 
atlas\footnote{see http://cfaku5.harvard.edu/}, and the chosen solar model 
atmosphere was that of Holweger \& M\"uller (1974)\nocite{HM74}.
The transition probabilities of the $n$-capture element transitions
were taken from the atomic physics literature without change (see 
the Appendix), but the $gf$'s of the surrounding atomic spectral 
features were altered to match the solar spectrum.
    Except for the syntheses described in \S3.2 for the determination
    of CNO abundances, the molecular lines were treated
    simply as contaminants.  Relative line strengths of different
    rotational lines of molecular electronic bands can be determined to
    better precision than overall band strengths. Therefore, for an
    individual synthesis we varied the absorptions of a given molecule
    as a set by simply varying the assumed abundance of C, N, or O.

However, the solar spectrum in the UV is dominated by overlapping 
saturated lines and there is no routinely accessible electronic
solar spectrum below $\lambda$~$\lesssim$ 3000~\AA.
Thus, following the procedure of Sneden et~al. (1998) we chose 
the bright UMP giant HD~122563 (4500~K, 1.30, --2.7, 2.5~\kmsec;
Westin et~al. 2000) as the template star for spectral regions covered 
by our HST-STIS spectrum.
This choice took advantage of the underabundance of $n$-capture elements
with Z~$\geq$ 56 in HD~122563, e.g., [Eu/Fe]~$\simeq$ --0.4 (Westin
et~al., and references therein).
This deficiency results in no detectable absorptions due to \third\
$n$-capture element transitions either in the HST-GHRS spectra of
Sneden et~al. or in our new HST-STIS spectrum of this star.
Thus HD~122563 provides an excellent template spectrum to assist
    in modeling the strengths of other atomic and molecular contaminants to
the $n$-capture transitions of interest.

After iterating the synthetic spectrum line lists to produce reasonable
agreement with the solar or HD~122563 spectrum, we then applied them 
to matching the spectrum of \bd17.
The resulting line-by-line \bd17\ $n$-capture abundances are presented 
in Table~\ref{tab2}. 
For unblended lines of $n$-capture species that are unaffected by 
isotopic or hyperfine structure problems, column~4 of Table~\ref{tab2}
gives the EW of the line that was used in the abundance derivation.
For the more complex features requiring synthetic spectrum computations, 
the notation ``syn'' is given in column~4.

The mean log~$\epsilon$ and [X/Fe] values for the $n$-capture elements 
in \bd17\ are given in Table~\ref{tab3}, along with sample standard 
deviations $\sigma$ and the number of lines analyzed.
The [X/Fe] ratios were computed by assuming that [Fe/H]~= --2.09, and
adopting the recommended solar abundances log~$\epsilon_{\sun}$ of
Grevesse \& Sauval (1998)\nocite{GS98}, which also are listed in 
Table~\ref{tab3}.
For a few elements, recent solar photospheric abundance studies
(e.g., Lawler et~al. 2001b for \ion{Tb}{2}) derive abundances that 
ultimately may supersede the solar numbers in Table~\ref{tab3}.
However, here we are using the [X/Fe] ratios only as general guides to
$n$-capture element enhancements, so for simplicity we will employ
only the Grevesse \& Sauval solar abundances.

For some elements with only a few detected lines, nominal $\sigma$ values 
turned out fortuitously to be very small.  
But the line-to-line scatters for elements possessing many 
transitions suggest that more realistic minimum measurement and 
internal analysis uncertainties for single lines are $\sim$0.05~dex.
Therefore we have adopted more conservative scatter values for
Table~\ref{tab3} and figures that use the data of this table:
for a species with three or more lines, $\sigma$ equals the greater of 
the measured value or 0.05~dex; for a species with two lines, the
greater of the measured value or 0.10~dex; and for a species with 
only one line, 0.15~dex.
The only exception to this is for U, which we consider below.

Two other recent spectroscopic studies have reported $n$-capture abundances
for \bd17.
Johnson \& Bolte (2001) derived abundances for seven rare-earth elements,
and the mean abundance difference is $<\Delta$[$n$-capture/Fe]$>$~$\equiv$
[$n$-capture/Fe]$_{\rm JB01}$~--~[$n$-capture/Fe]$_{\rm us}$~=
+0.02~$\pm$~0.04 ($\sigma$~= 0.10).
This is excellent agreement considering that much of the
improved atomic transition data had not been published at the time of
Johnson \& Bolte's analysis, and that their results were based on far fewer
lines per element than we have used.
The comparison with Burris et~al.  (2000) for eight $n$-capture elements
seems at first glance to be much worse: $<\Delta$[$n$-capture/Fe]$>$~=
+0.25~$\pm$~0.06 ($\sigma$~= 0.18).
However, the spectroscopic resolution for that study (R~$\simeq$ 30,000)
was much less than the resolutions of our Keck and McDonald data, making
fewer transitions available for analysis.
Burris et~al. determined abundances from only 1--2 lines per element,
and assumed a larger microturbulent velocity (\vt~= 2.3~\kmsec) than
we recommend (\vt~= 1.9~\kmsec).
The \vt\ choice strongly affects the abundances derived from \ion{Sr}{2}
and \ion{Ba}{2} lines.
If we neglect those two elements and also Nd, whose abundance is subject
to large uncertainties from its atomic transition data, then the
comparison is much better for the other remaining five elements in
common: $<\Delta$[$n$-capture/Fe]$>$~= +0.14~$\pm$~0.04 ($\sigma$~= 0.09).

\begin{figure}
\epsscale{0.95}
\plotone{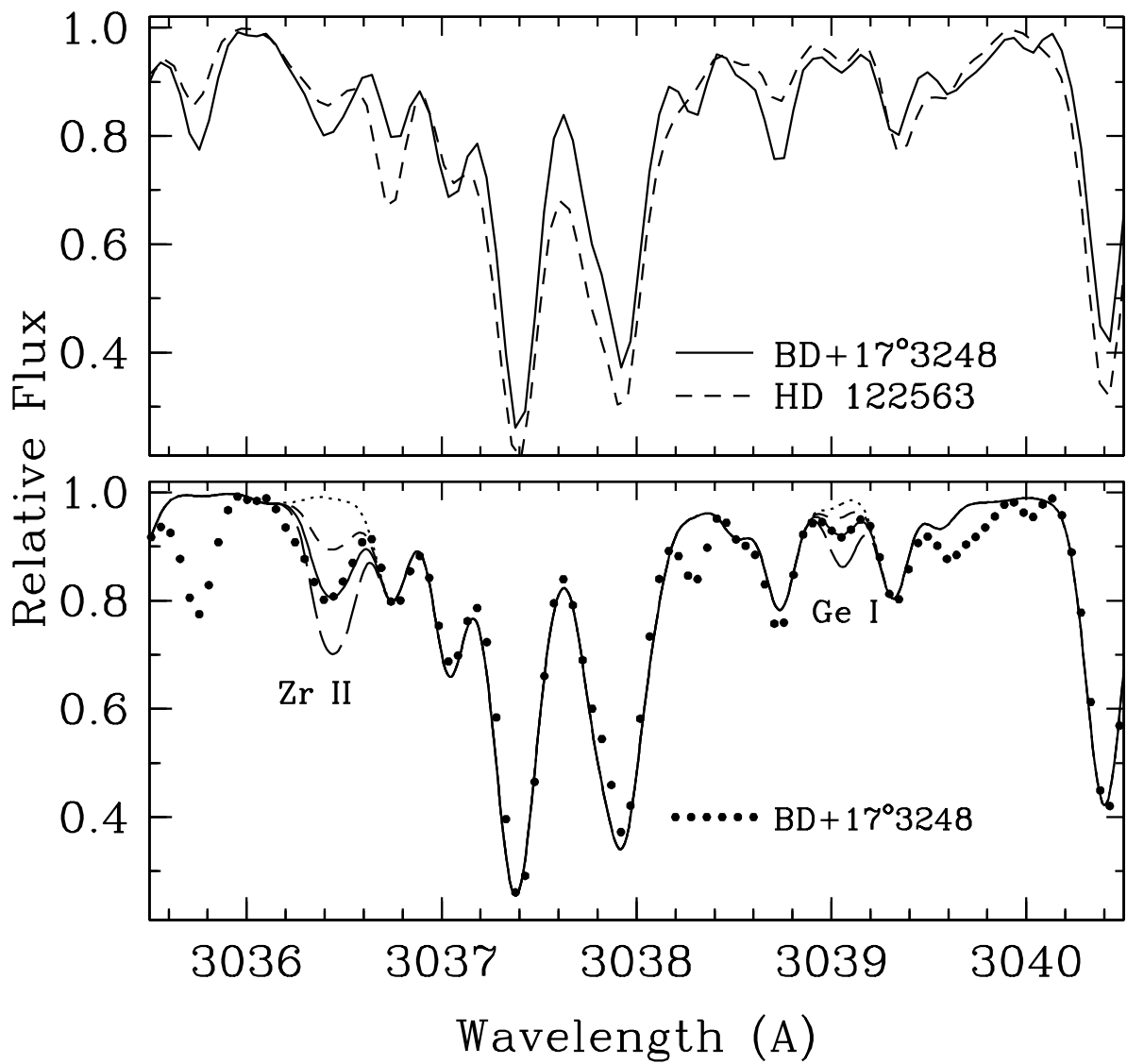}
\caption{ 
Observed HST-STIS and synthetic spectra in region surrounding the
3036.39~\AA\ \ion{Zr}{2}, 3036.52~\AA\ \ion{Zr}{2}, and 3039.07~\AA\ 
\ion{Ge}{1} lines.
In the top panel, the observed \bd17\ spectrum is compared to that
of HD~122563.
In the bottom panel, the observed \bd17\ spectrum is compared to four
synthetic spectra, shown in order of increasing abundance of Zr and Ge
by dotted, short dashed, solid, and long dashed lines computed for these
abundances: log~$\epsilon$(Zr)~= --$\infty$, +0.30, +0.70, +1.10,
and log~$\epsilon$(Ge)~= --$\infty$, +0.20, +0.60, +1.00.
\label{f3}}
\end{figure}

\begin{figure}
\epsscale{0.95}
\plotone{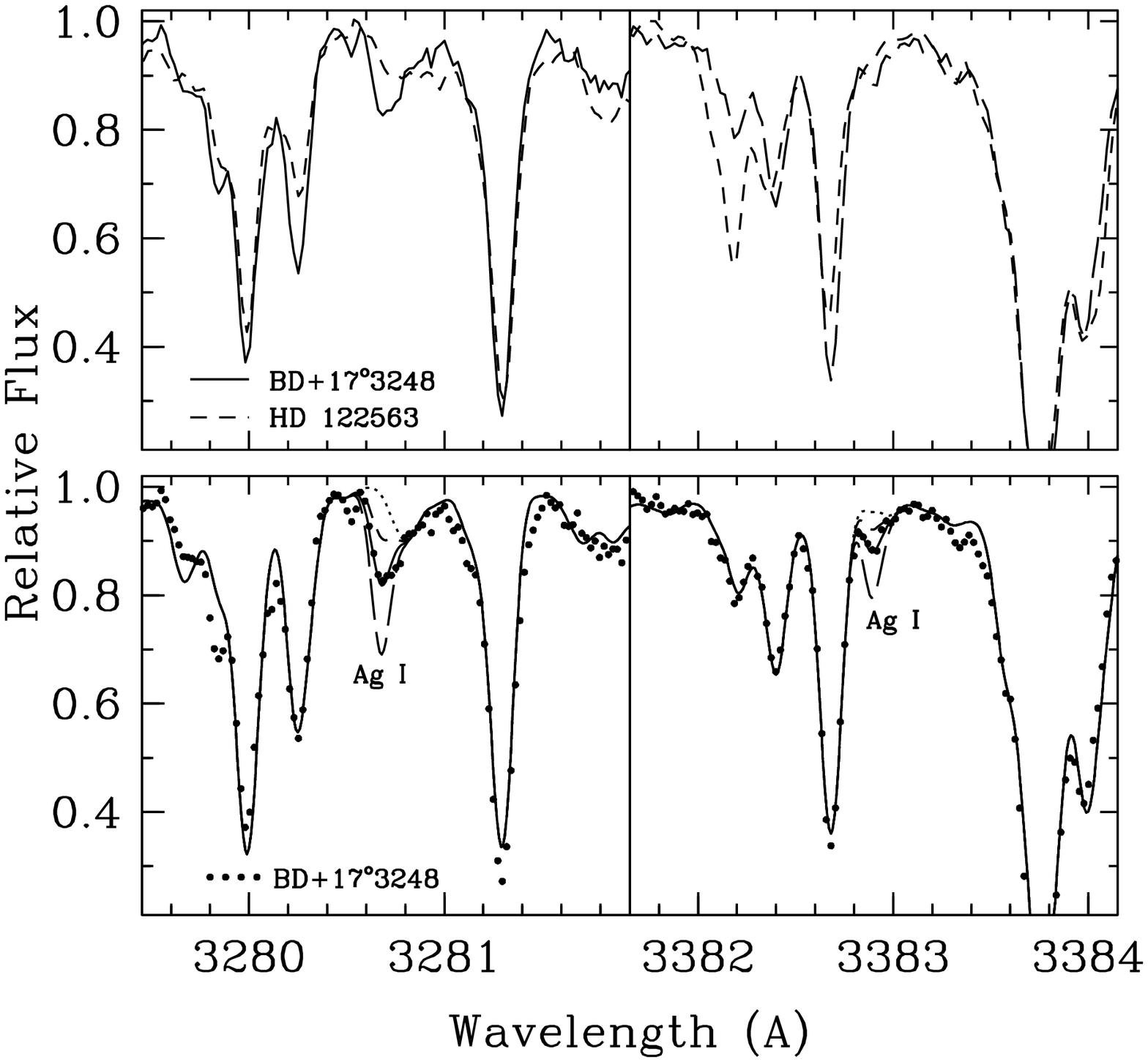}
\caption{
Observed Keck-HIRES and synthetic spectra in region surrounding the
\ion{Ag}{1} 3280.67 and 3382.90 \AA\ lines, in the manner of
Figure~\ref{f3}.
In the bottom panel, the synthetic spectra were computed for
abundances log~$\epsilon$(Ag)~= --$\infty$, --0.68, --0.28, +0.12.
\label{f4}}
\end{figure}

\begin{figure}
\epsscale{0.95}
\plotone{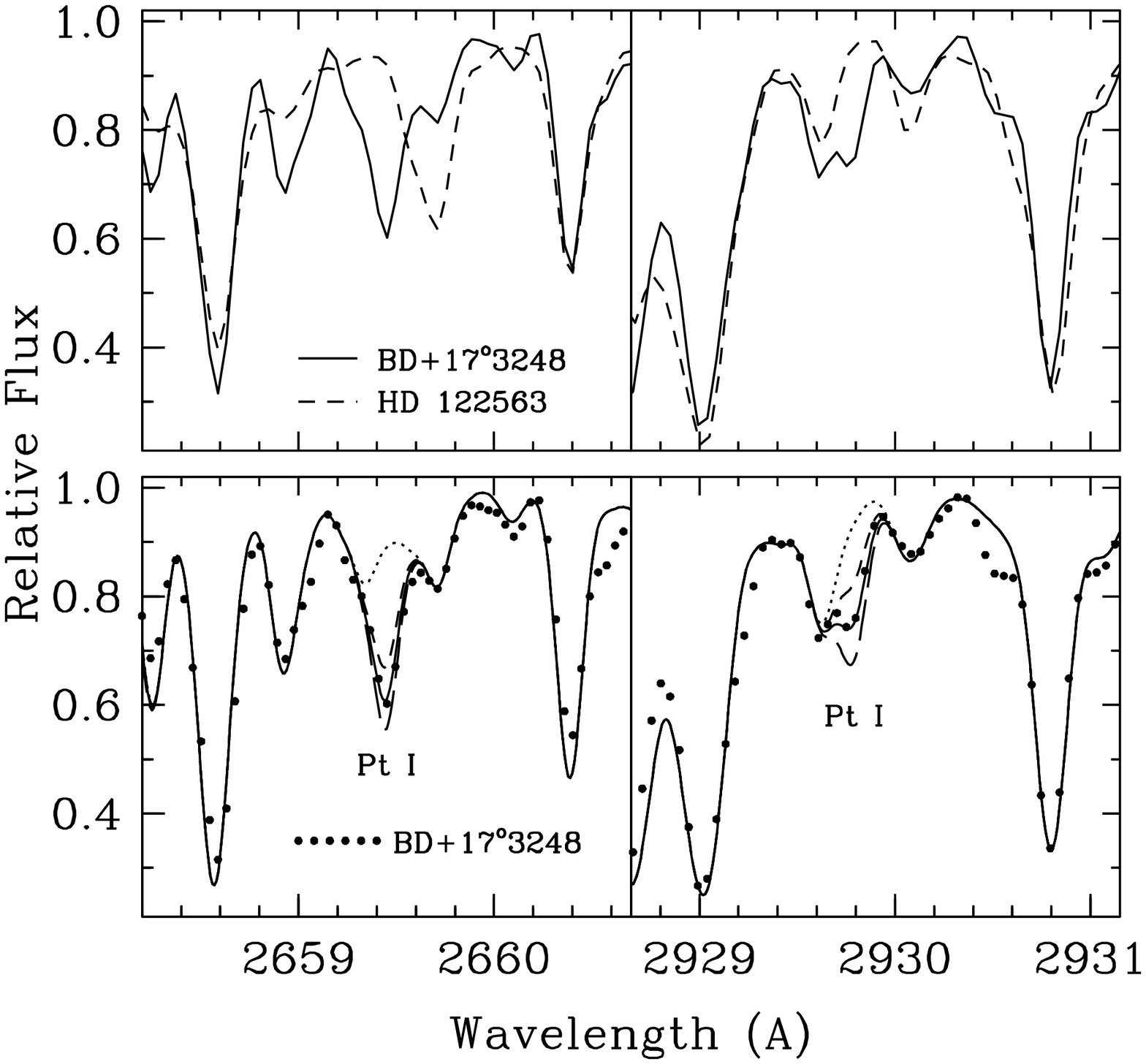}
\caption{
Observed HST-STIS and synthetic spectra in region surrounding the
\ion{Pt}{1} 2659.45, 2929.78 \AA\ lines, in the manner of
Figure~\ref{f3}.
In the bottom panel, the synthetic spectra were computed for
abundances log~$\epsilon$(Pt)~= --$\infty$, +0.27, +0.67, +1.07.
\label{f5}}
\end{figure}

\begin{figure}
\epsscale{0.95}
\plotone{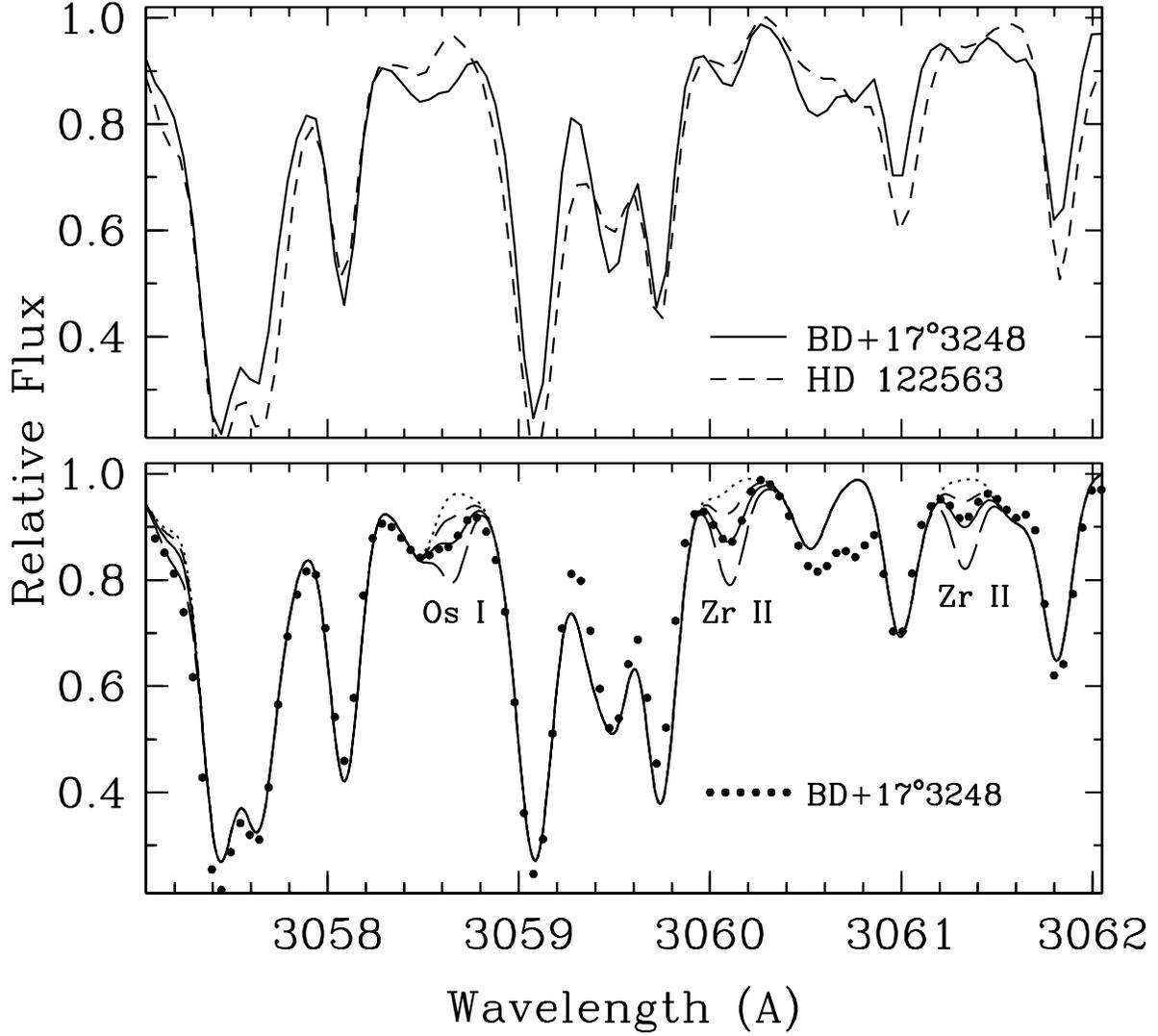}
\caption{
Observed HST-STIS and synthetic spectra in region surrounding the
3060.11~\AA\ \ion{Zr}{2}, 3061.33~\AA\ \ion{Zr}{2}, and 3058.66~\AA\
\ion{Os}{1} lines, in the manner of Figure~\ref{f3}.
In the bottom panel, the synthetic spectra were computed for
abundances log~$\epsilon$(Zr)~= --$\infty$, +0.30, +0.70, +1.10,
and log~$\epsilon$(Os)~= --$\infty$, +0.05, +0.45, +0.95.
\label{f6}}
\end{figure}

\begin{figure}
\epsscale{0.95}
\plotone{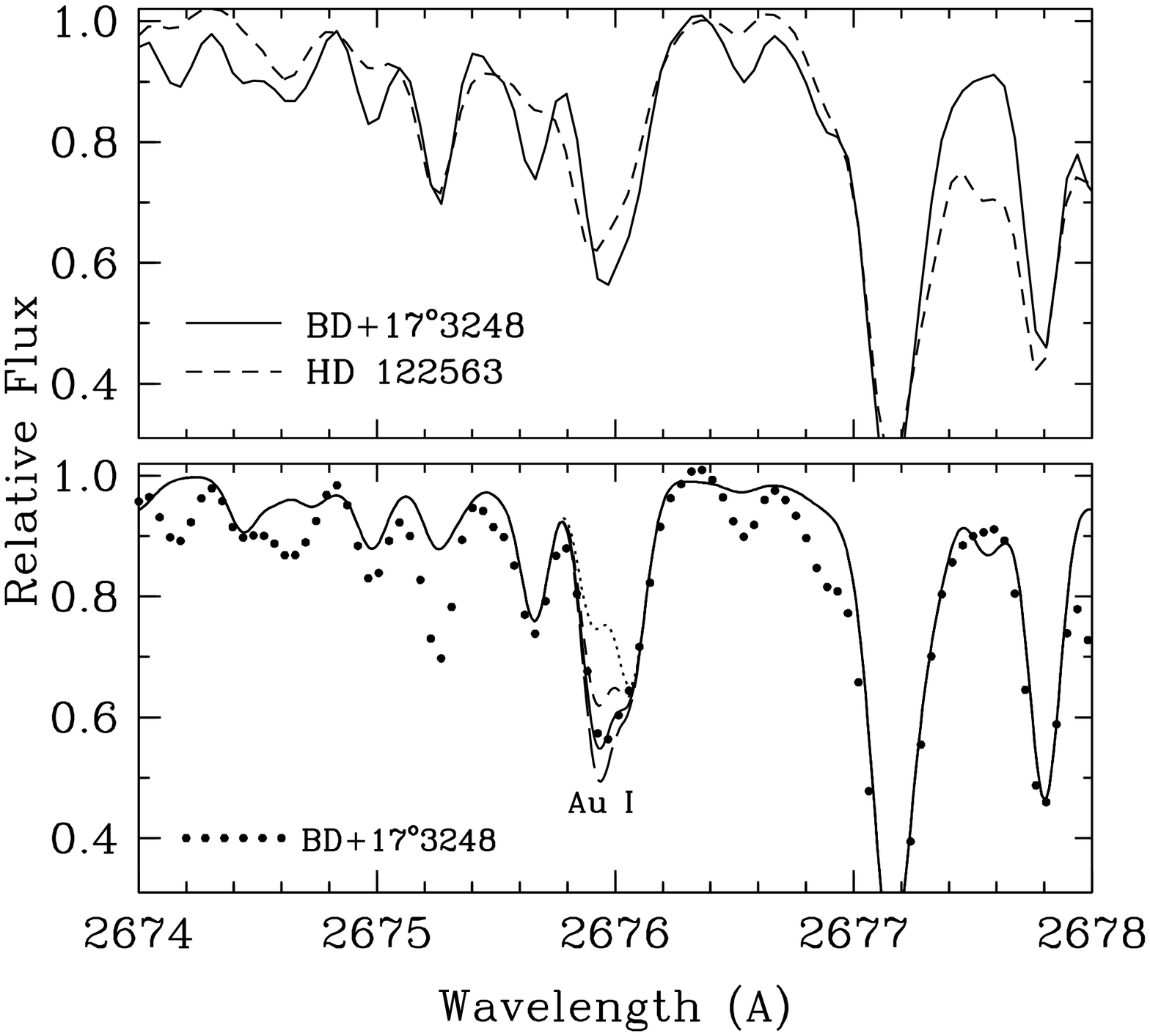}
\caption{
Observed HST-STIS and synthetic spectra in region surrounding the
\ion{Au}{1} 2675.94~\AA\ line, in the manner of Figure~\ref{f3}.
In the bottom panel, the synthetic spectra were computed for abundances
log~$\epsilon$(Au)~= --$\infty$, --0.82, --0.32, +0.18.
\label{f7}}
\end{figure}

\begin{figure}
\epsscale{0.95}
\plotone{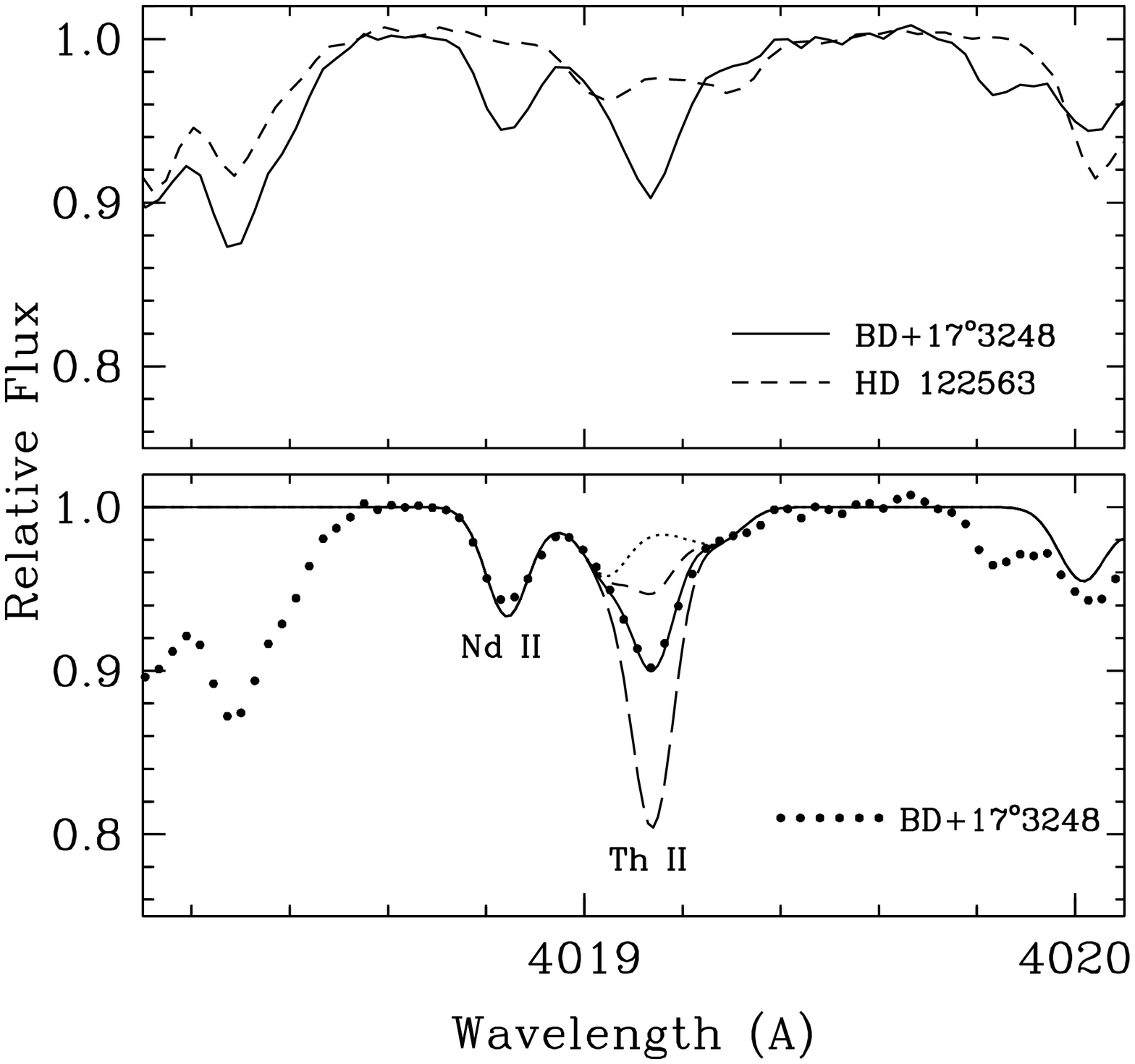}
\caption{
Observed Keck-HIRES and synthetic spectra in region surrounding the
\ion{Th}{2} 4019.12 \AA\ line, in the manner of Figure~\ref{f3}.
In the bottom panel, the synthetic spectra were computed for
abundances log~$\epsilon$(Th)~= --$\infty$, --1.58, --1.18, --0.78.
\label{f8}}
\end{figure}

In Figures~\ref{f3} to~\ref{f8} we present a series of small
spectral regions surrounding some key $n$-capture transitions.
The top panels of each of these figures compare the observed spectra
of \bd17\ to that of HD~122563, and the bottom panels match synthetic
and observed spectra just of \bd17.
Inspection of these figures reveals that the line strengths of the two 
stars generally are comparable, as the $\sim$700~K cooler \teff\ of
HD~122563 strengthens its absorption spectrum enough to roughly
compensate for its $\sim$0.6~dex lower metallicity.
In a few instances an atomic or molecular line is stronger in the 
HD~122563 spectrum than it is in \bd17, due to a particular combination 
of excitation/ionization conditions of the transition.
But far more often  some lines are  much stronger
in the \bd17\ spectrum; these are the signs of very large
abundances of most $n$-capture elements in this star.

Some comments should be given here on general $n$-capture abundance results.
Consider first ``rare-earth'' elements, defined as those of the lanthanide 
group (La$\rightarrow$Lu, 57~$\leq$ Z~$\leq$ 71) plus Ba (Z~= 56).
They are linked not only by consecutive atomic numbers in the 
Periodic Table but also by spectroscopic species availability. 
Rare-earths have low first-ionization potentials (I.P.~$\leq$ 6.2~eV), so
these elements exist almost entirely as singly ionized species in cool 
metal-poor giants: N$_{X tot}$~$\simeq$ N$_{X II}$. 
Thus the ``Saha corrections'' for other ionization stages are negligible.
Additionally, the detectable transitions arise predominantly from low 
lying excitation states (usually E.P.~$<$ 1.0~eV) of these ions.
Therefore, the Boltzmann exponential factors essentially cancel out for
abundance ratios among the rare-earth elements.
Derived ratios such as, e.g., log~$\epsilon$(La/Dy), are nearly 
independent of \teff\ and log~$g$ uncertainties in \bd17.

The mean of the measured line-to-line scatters $\sigma$ of the rare-earth 
elements given in Table~\ref{tab3} (ignoring the $\sigma$ values
arbitrarily defined for Pr, Tb, and Tm) is not large: $<\sigma>$~= 0.10. 
It would reduce to 0.08 if we neglect the values for Nd (see
    the discussion in the appendix) and Ho (whose abundance is
    based on three lines, two of which are badly blended).
Undoubtedly the mean $\sigma$ would shrink further with careful reexamination
of some obvious outlying abundance points for a few elements.
The insensitivity of rare-earth element abundance ratios and the
availability of excellent atomic data for these elements combine to
yield a very accurate ``complete'' abundance distribution of rare-earths 
in \bd17.
The entire set of these elements is greatly enhanced, reaching a maximum of
[$n$-capture/Fe]~$\simeq$ +0.9 for atomic numbers Z~$>$ 60.

The abundances of lighter $n$-capture elements (Ge$\rightarrow$Ag,
32~$\leq$ Z~$\leq$ 71) are less overabundant, and the
abundance enhancements apparently vary directly with atomic number.
The lightest of these elements, Ge, is in fact very $underabundant$
in \bd17: [Ge/Fe]~$\sim$ --0.7.
This effect can be seen in Figure~\ref{f3}, which in the top
    panel overlays the observed spectra of \bd17\ and HD~122563
    surrounding the \ion{Ge}{1} 3039.07~\AA\ line, and in the bottom
    panel shows a comparison between synthetic and observed spectra
    of that wavelength region in \bd17.
If [Ge/Fe] were actually $\sim$0 then the \ion{Ge}{1} line would be
stronger, not weaker, than the two lines bracketing it 0.3~\AA\ to the
blue and to the red.
 
The abundances of the Sr$\rightarrow$Zr group (38~$\leq$ Z~$\leq$ 40)
are at best weakly overabundant with respect to Fe (this is similar to
some other $r$-process-rich stars, such as HD~115444; see e.g., Figures~4
and 5 of Westin et~al. 2000).
Both \ion{Sr}{1} and \ion{Sr}{2} lines have been analyzed, and neither
species indicates an anomalously large abundance\footnote{
The abundance derived from the single detected \ion{Sr}{1} line is
0.2~dex smaller than that from the two \ion{Sr}{2} lines.
However, the ionized lines are the two extremely strong resonance
transitions on the flat/damping region of the curve-of-growth
(log(EW/$\lambda$)~$\simeq$ --4.4) and Sr abundances derived from
    them are very sensitive to the adopted \vt\ value.
The error bars of Sr abundances derived from neutral and ionized
lines overlap.}.
Additionally, the very modest positive [Zr/Fe] ratio derived from 19
\ion{Zr}{2} transitions appearing on our ground-based spectra is confirmed
by five lines of this species appearing on our HST-STIS spectrum; see some
of the UV \ion{Zr}{2} lines in Figures~\ref{f3}~and~\ref{f6}.
 
Elements in the Z~= 41--47 range are more overabundant with respect to Fe.
Synthetic and observed spectra of the two \ion{Ag}{1} resonance lines
are presented in Figure~\ref{f4}, revealing cleanly detected features
that are relatively easy to analyze.
The \ion{Ag}{1} lines are the resonance transitions, which have well-known
$gf$ values and hyperfine/isotopic structure information.
The major surrounding contaminant atomic and NH molecular features
can be modeled to good approximation.
Therefore the derived Ag abundance should be reliable.
 
Detected transitions of \third\ $n$-capture peak elements 
(Os$\rightarrow$Pb, 76~$\leq$ Z~$\leq$ 82) have several 
common properties: they all arise from low-excitation states of
    the neutral species; they generally occur at very short wavelengths,
    $\lambda$~$\lesssim$ 3800~\AA; and they are mostly weak lines that suffer
    blending from other atomic and molecular (mostly OH and NH) features.
Fortunately, transition probabilities for these lines are generally
known to good accuracy, and the abundance analysis limitation lies
mostly with line detection and modeling of contaminant spectral
features.
Some of the details involved in line selection for these elements
have been discussed by Sneden et~al. (1998);  the Appendix 
provides additional information.

The HST-STIS spectra of two \ion{Pt}{1} lines are displayed in
    Figure~\ref{f5} and one of the \ion{Os}{1} lines in
    Figure~\ref{f6}.
The top panels of these two figures show very clearly the presence of
absorption lines of \third\ $n$-capture-peak species in \bd17\
and their undetectability in HD~122563.
The fits of synthetic to the observed spectra of \bd17\ displayed
in the bottom panels confirm the conclusions of the visual 
comparison of the two stars.

Abundances of \third\ $n$-capture peak elements Os, Ir, Pt, and Au in \bd17\
are very consistent: [$<$Os,Ir,Pt,Au$>$/Fe]~= +0.99~$\pm$~0.07 
($\sigma$~= 0.14).
This mean overabundance is slightly larger than that of the 
rare-earth elements whose solar-system abundances are dominated by the
    $r$-process: [$<$Eu$\rightarrow$Tm$>$/Fe]~= +0.88~$\pm$~0.03
    ($\sigma$~= 0.07), but the difference is hardly more than the
    combined standard deviations.
This $\sim$0.1~dex offset may be a real abundance effect, 
but it possibly could be simply an uncertainty
in our abundance computations.  
Recall that the \third\ $n$-capture-peak
 abundances have been derived from neutral-species 
transitions, while those of the rare-earths have been derived
from ionized-species lines. 
Thus, uncertainties in \teff\ and log~$g$ enter into ionization 
equilibrium computations.
While the estimated uncertainties of $\pm$150~K in \teff\ and
$\pm$0.3 in log~$g$ produce no changes larger than $\simeq$0.02 in
abundance ratios log~$\epsilon$(rare-earth/rare-earth) or
log~$\epsilon$(\third\-peak/\third\-peak), the temperature uncertainty
yields an uncertainty of $\mp$0.08 in
log~$\epsilon$(\third\-peak/rare-earth), and a gravity uncertainty
leads to $\pm$0.12 (see also \S 3.5 of SMPCBA and \S 3.1 of Westin et~al.
2000).  Therefore a small increase in \teff\ or a small decrease 
in log~$g$ could make the $\sim$0.1 offset between the abundances of
\third\-peak and rare-earth elements disappear.
Note that the slightly higher abundances of \third\-peak elements 
with respect to the rare-earths makes it unlikely that departures
from LTE in the form of over-ionization is causing the small offset
between the two element groups.\footnote{
The first-ionization potentials of Os, Ir, Pt, and Au are larger
than those of most other metals, averaging $\sim$8.9~eV.
Radiative ionization of these species would require very energetic
UV photons, which are not in plentiful supply in our program stars.}

We have detected gold in the STIS spectrum of \bd17.
In the top panel of Figure~\ref{f7} we plot observed spectra of the 
2675.94~\AA\ \ion{Au}{1} resonance line \bd17\ and HD~122563, and in
the bottom panel of this figure we compare observed and synthetic spectra
of \bd17\ alone.  
There is little obvious difference between the spectra of \bd17\ and 
HD~122563 at 2676.0~\AA, but we believe that this is due to the 
complex set of absorption lines that contribute to the total feature 
at this wavelength.
The 2675.94~\AA\ \ion{Au}{1} line is blended by 2675.89~\AA\ OH,
2675.90~\AA\ \ion{Ta}{2}, 2675.94~\AA\ \ion{Nb}{2}, 2676.07~\AA\ \ion{Ti}{1}, 
and 2676.09~\AA\ \ion{Mn}{1}.
In particular, OH has strong absorption lines throughout the UV spectrum 
of HD~122563, leading in this case to a deep 2676.0~\AA\ feature.  
In \bd17\ the OH absorption does not disappear but it is much less prominent.  
To determine an abundance for Au we must also deal with the other 
atomic contaminants.  
We did not determine an abundance for Ta, but Burris et~al. (2000)
suggest that the $r$-process fraction of this element in solar system
material is 0.60.
This roughly equal split between $r$- and $s$-process contributions
is similar to those of rare-earths Ce, Pr, and Nd, and so we estimated
for \bd17\ that log~$\epsilon$(Ta)~$\simeq$
 log~$\epsilon$(Ta)$_{\sun}$~--~[Fe/H]~+~[$<$Ce,Pr,Nd$>$/Fe] $\simeq$
+0.13~--~2.09~+~0.6 $\simeq$ --1.4.
The resulting Ta and Nb absorptions account for about half of the
total feature strength at 2675.9~\AA.
The derived Au abundance is thus vulnerable to abundance or
transition probability uncertainties in Ta and Nb.
Comments on the other \ion{Au}{1} resonance line at 2427.94~\AA\
appear in the Appendix.

The \third\ $n$-capture peak element Pb merits an additional comment.
We are unable to detect the \ion{Pb}{1} lines at 3683.46 and 4057.81~\AA\
in our ground-based spectra of \bd17.
The 2833.05~\AA\ line should be much stronger, and indeed there
is detectable absorption near this wavelength in the HST-STIS spectrum.
Frustratingly, the observed line centroid lies about 0.03~\AA\ redward
of the predicted \ion{Pb}{1} line, and there is an \ion{Fe}{2}
line with a known laboratory $gf$ that appears to account for the
majority of the strength of the combined feature.  
We have listed a rough estimate of the Pb abundance from this line
    in Table~\ref{tab2}, but the upper limits derived from the
    longer wavelength lines provide just as reliable a Pb abundance;
    hence we suggest 
in Table~\ref{tab3} that log~$\epsilon$(Pb)~$\lesssim$ +0.3.

The well-studied \ion{Th}{2} 4019.12~\AA\ line is easily detected in \bd17, 
while in HD~122563 the \ion{Th}{2} line is lost amid other absorbers 
at this wavelength; see the top panel of Figure~\ref{f8}.
Details of the analysis of this line have been discussed in several
studies (Morell, K{\"a}llander, \& Butcher 1992\nocite{MKB92};
Fran\c{c}ois, Spite, \& Spite 1993\nocite{FSS93}; SMPCBA;
Norris et~al. 1997\nocite{NRB97}; Westin et~al. 2000; 
Johnson \& Bolte 2001). 
We can add little to the discussion in those papers, but emphasize from
inspection of the synthetic spectra given in the lower panel of 
Figure~\ref{f8} that \bd17\ has a fairly clean \ion{Th}{2} feature,
and the derived Th abundance is not very sensitive to the modeling
of the contaminants.
The ratio of Th to the rare-earth elements is determined to high accuracy.

\begin{figure}
\epsscale{0.95}
\plotone{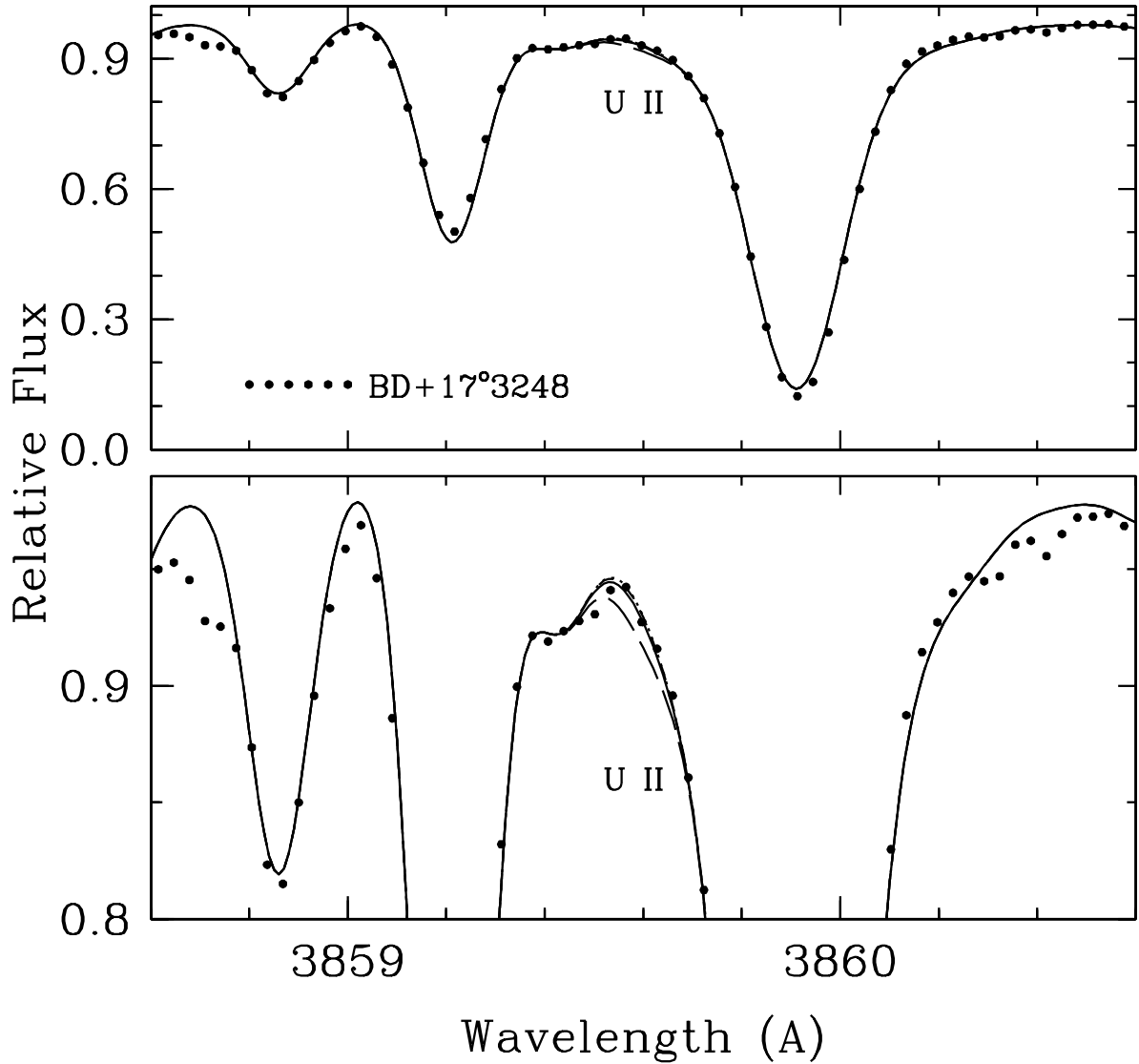}
\caption{
Spectra of \bd17\ in region surrounding the \ion{U}{2} 3859.57 \AA\ line.
In both panels the observed Keck-HIRES spectrum is compared to four
synthetic spectra computed for abundances log~$\epsilon$(U)~= 
--$\infty$, --2.60, --2.10, --1.60. (Note that some of the lines 
overlap.)
The only difference between the two panels is the vertical scale, which
is expanded in the bottom panel for easier assessment of the extremely
weak \ion{U}{2} feature.
\label{f9}}
\end{figure}

We also suggest from syntheses of the 3859.60~\AA\ \ion{U}{2} line 
(Figure~\ref{f9}) that it is possibly detected in \bd17.
The line is of course extremely weak, and to maximize its visibility we
have experimented with spectral deconvolution of the Keck-HIRES spectrum
in this wavelength region.
The deconvolution was done according to the algorithm of Gilliland
et~al. (1992)\nocite{GMWEL92}, with the purpose of restoring the true 
spectral resolution limit of the data.
Such an operation of course decreases the S/N per pixel as it increases
the resolution.
The observed spectrum in Figure~\ref{f9} is the result of the 
deconvolution process, but the very weak \ion{U}{2} line appears
to be present in both the original and deconvolved spectrum of \bd17.
Naturally, the derived U abundance has an associated error estimate that
is much larger than that of most other $n$-capture elements.

\section{DISCUSSION}

\subsection {Neutron-Capture Element Abundances}

Our Keck and HST observations have provided abundance determinations 
for a wide range in proton number,
from Ge to U. This is currently the largest range of $n$-capture elements
detected in any metal-poor halo stars. 

\subsubsection{Heavier Elements (Z $\ge$ 56)}

\begin{figure}
\epsscale{0.95}
\plotone{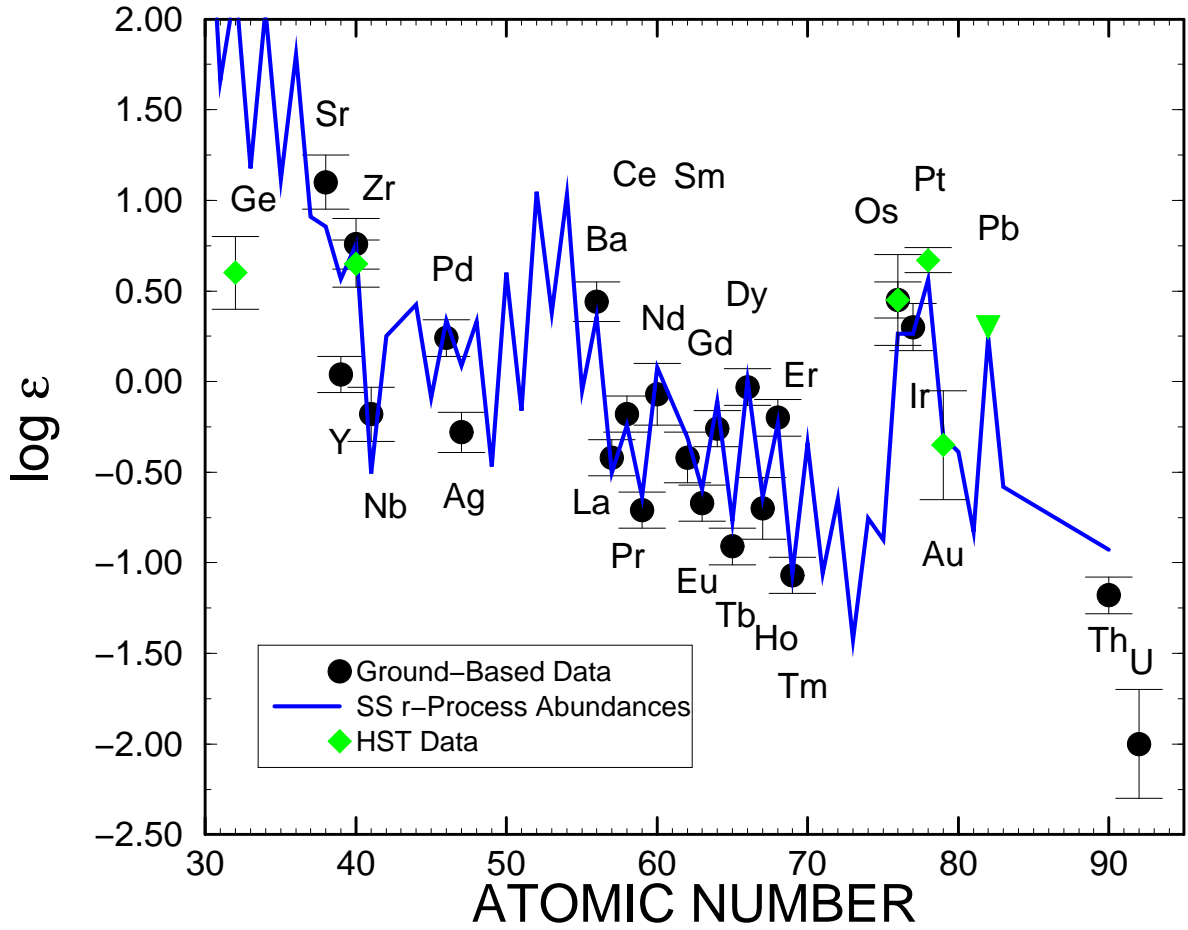}
\caption{
Neutron-capture element abundances in \bd17, obtained by ground-based 
and HST observations, compared to a scaled solar system $r$-process 
abundance curve.
The upper limit on the lead abundance is denoted by an inverted triangle.
Note also the thorium  and uranium detections.
\label{f10}}
\end{figure}

Abundance determinations in a number of Galactic halo and 
globular cluster stars have shown that the
rare-earth elements, roughly from barium to hafnium, are consistent with a
scaled (i.e., adjusted for metallicity) solar system $r$-process 
distribution
(see e.g., SMPCBA; Cowan et~al. 1999; BPASCR; Sneden et~al 2000a, 2000b; 
Johnson 
\& Bolte 2001\nocite{sne96,cow99,bur00,sne00a,sne00b,jb01}). 
Isotopic abundances for europium in metal-poor halo stars also 
indicate a solar system pattern, as shown in Sneden et~al. 2001b.
Examination of  Figure~\ref{f10}, 
where the solid line denotes the solar 
system $r$-process-only elemental abundances,   
again shows that same pattern.
What is different about this star, however, is that the HST observations
demonstrate that the \third\ $r$-process peak elemental abundances
(now including gold and the upper limit on lead) 
appear to follow the rare-earth pattern and
fall on the same matching abundance distribution. 
There is some indication, however, that a few  of the elements, for example Os,
in the \third\ $r$-process  
peak may be slightly higher (on the order of 0.1 dex, which is the typical
elemental abundance uncertainty) than the matching
solar 
system $r$-process curve. As noted earlier in \S 3.3, this small offset
may be a real effect or may result from uncertainties in our abundance
computations - the difference between abundances of neutral and 
ionized species. 
We note further that while Os appears somewhat high,
Ir and Au fall precisely on the curve, 
and that seems to point to an overall agreement with the rare-earth region.
There have been suggestions  of this agreement 
before in the observations of 
HD 115444 (Westin et~al. 2000\nocite{wes00}) and CS~22892-052 (Sneden
et~al. 2000a\nocite{sne00a}), but these new combined (ground- and 
space-based) observations of \bd17\ have provided greater access to 
the heaviest 
stable  $n$-capture elements than previously possible,
with  presumably more reliable abundance determinations. 

\begin{figure}
\plotone{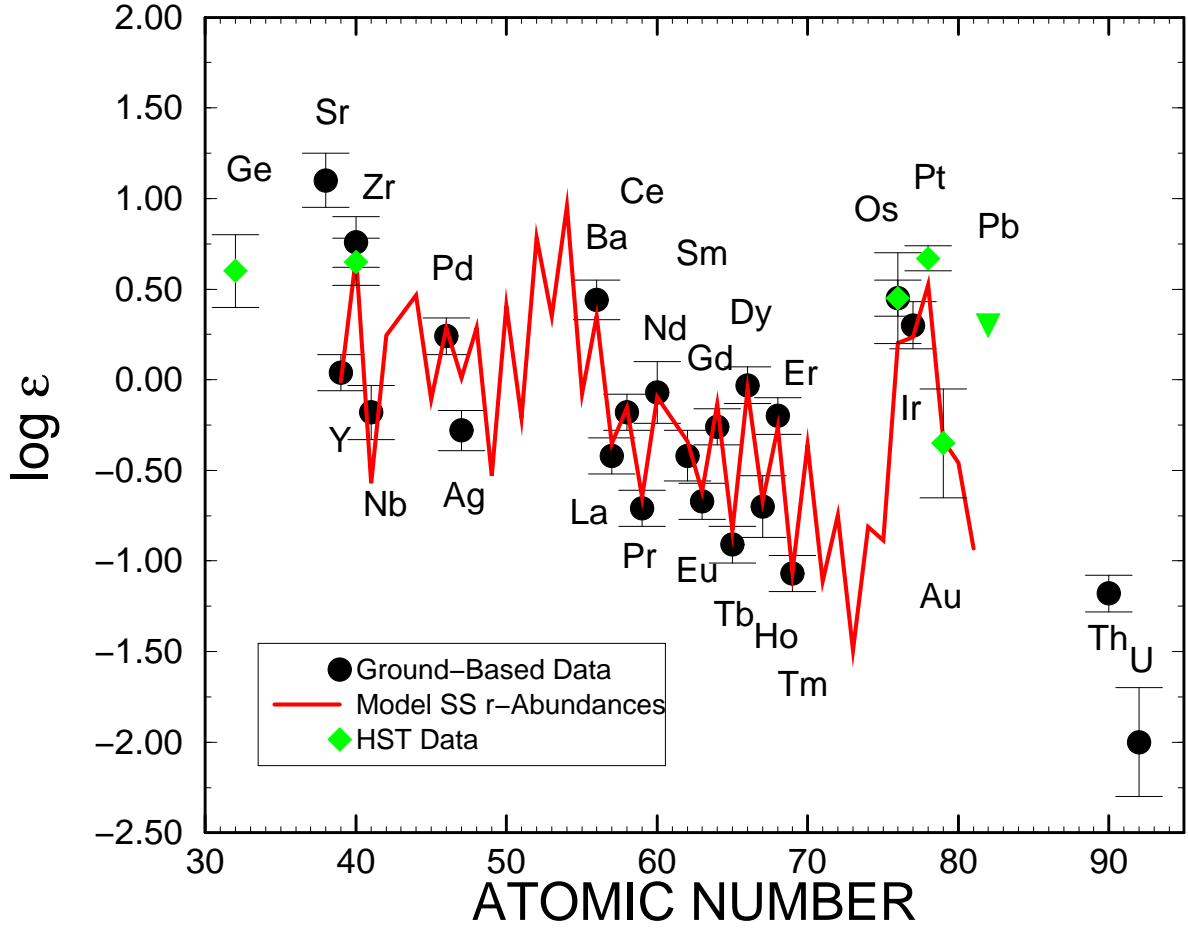}
\caption{
Same as Figure~\ref{f10} except the scaled solar system $r$-process curve
is based upon the stellar $s$-process model of Arlandini et al. (1999).
\label{f11}}
\end{figure}

The solid-line curve illustrated in Figure~\ref{f10} was derived 
by summing the individual isotopic contributions from the {\it s}- and the
{\it r}-process in solar system material, as determined by
K{\"a}ppeler et~al. (1989)\nocite{kap89} and Wisshak, Voss, \& 
K{\"a}ppeler  
(1996)\nocite{wis96} from $n$-capture cross section
measurements. 
This solar $r$-process, and corresponding $s$-process, elemental distribution
was first published in SMPCBA and later updated in 
BPASCR. 
The deconvolution of the solar system material into the $s$-process and 
$r$-process relied upon reproducing the ``$\sigma$ N'' curve ( i.e., the 
product of the $n$-capture cross-section and $s$-process abundance).
This ``classical approach'' to the $s$-process is
empirical and by definition  model-independent. More sophisticated 
models, based upon $s$-process nucleosynthesis  in low-mass AGB stars,   
have been developed recently. In  Figure~\ref{f11} we again compare the 
abundances of elements in \bd17\ with a solar $r$-process curve. But in
this case the curve has been obtained by summing the solar $r$-process 
isotopic residuals from
the stellar $s$-process model of Arlandini et~al. (1999\nocite{arl99}).
(Note that Sr and Pb are not included in this distribution.)
As is seen in this figure,
the agreement between this ``model'' solar system
$r$-process curve and the rare-earth abundances in \bd17\ is 
slightly  better
than the fit of stellar abundances to the standard $r$-process curve 
shown in Figure~\ref{f10}.
For the 13 elements with 56~$\leq$ Z~$\leq$ 69, the mean abundance
comparison with the Arlandini et~al. $r$-process solar distribution
is $<\Delta$(log~$\epsilon)>$~$\equiv$ 
$<$log~$\epsilon_{BD~+17^{\circ}3248}$--log~$\epsilon_{solar}>$~= +0.395~$\pm$~0.017 ($\sigma$~= 0.062).
In the comparison of \bd17\ abundances with the standard solar $r$-process 
curve the mean difference is nearly the same but the scatter
of individual points is somewhat larger:
$<\Delta$(log~$\epsilon)>$~= +0.408~$\pm$~0.024 ($\sigma$~= 0.082).

The agreement between the heavier $n$-capture  elements and the scaled
solar system $r$-process curve 
in this star, as well as in several  other stars,
suggests that the $r$-process is robust for the upper  end.
This in turn puts constraints on possible sites for the $r$-process -- something
that is not known with certainty --  
and on the conditions under which it operates. 
This robustness, for example, might suggest a narrow range of supernovae masses
(e.g., Mathews, Bazan, \& Cowan   1992; Wheeler, Cowan, \& Hillebrandt
1998) or 
perhaps well-confined conditions for 
$r$-process nucleosynthesis in a  supernova environment (e.g., Cameron 2001).

In Figures \ref{f10} and \ref{f11} it can be seen that
Ba (an element that is usually thought of as an {\it s}-process element and
is predominantly produced by that process in solar system material)
was, instead, synthesized by the $r$-process
in BD +17\deg 3248.  
These results support earlier suggestions (Truran 1981)\nocite{tru81} 
and the results of larger surveys (McWilliam et~al. 1995a,b; 
McWilliam 1998; BPASCR) 
of an $r$-process origin for this element early 
in the history of the Galaxy. 
It is also evident 
for \bd17\ that not only Ba, but all the other so-called $s$-process 
elements appear to have been synthesized early in the history of the 
Galaxy by the $r$-process, a result similar to that found for the 
individual UMP stars CS~22892-052 and HD~115444. 
The presence  of this 
{\it r}-process material in metal-poor stars such as \bd17\ and old halo stars,
including the heaviest {\it r}-process  elements,
also places constraints on the earliest stellar generations,
in particular,  suggesting
rapidly evolving progenitors of the halo stars.
This follows since the early Galaxy appears to be chemically inhomogeneous
in $n$-capture elements
(BPASCR; Sneden et~al. 2001; Cowan, Sneden, \& Truran  
2001)\nocite{bur00,sne01a,cow01} implying a relatively
short (with respect to stellar evolutionary) time scale between the
death of the progenitors and
the formation of the halo stars.
As an example of the scatter seen in the early Galaxy at low metallicity,
\bd17\ with   
[Fe/H] = --2.0 has a value of [Eu/Fe] = 0.85, 
significantly above the solar value.

\subsubsection{Lighter  Elements (30 $<$ Z $<$ 55)}

Our new observations of \bd17\ have demonstrated  the presence of 
Nb, Pd and Ag. 
This is the second metal-poor halo star, after CS~22892--052, with derived
abundances for several elements in the domain 40~$<$~Z~$<$~56, along
with abundances of nearly all rare-earth elements.
Examination of   Figure~\ref{f10} indicates that  the abundances of 
these three lighter $n$-capture are not consistent with the same solar
$r$-process curve that matches the heavier $n$-capture elements, i.e., Z $\ge$ 56.
These  observations of \bd17, along with those in CS~22892--052,
are consistent with earlier suggestions 
 -- based upon analyses of solar system meteoritic material --
that there may be two
sites for the {\em r}-process
with the heavier elements (Ba and above)
produced in more rapidly occurring events and the lighter elements in
less commonly occurring syntheses (Wasserburg, Busso, \& Gallino 1996;
Wasserburg \& Qian 2000). 
Alternatively, a combination of $r$-process nucleosynthesis in supernovae and 
in neutron star binaries (Rosswog et~al. 1999)\nocite{ros99}, with different 
sites responsible for different ends of the abundance distribution, 
is also possible.
Sneden et~al. (2000a) suggested that 
the total $n$-capture abundance pattern in CS~22892-052,
similar to what  we now find in \bd17, could also be consistent with
a neutrino-heated supernova ejecta $r$-process in a single supernova event,
with two different epochs in the explosion/ejection process. 
Some studies (see e.g., Pfeiffer et~al. 2001a; Pfeiffer, Ott, \& Kratz 2001b) 
were able to reproduce the 
total abundance pattern found in CS~22892-052 in the context of a single 
site.
We emphasize that the heavier $n$-capture elements -- that 
follow the scaled solar system $r$-process curve - all seem to 
be produced in one site, or at least under  similar conditions,
making this a robust process.  Thus,  for those
elements a second $r$-process site  (such as merging neutron-star 
binaries) is not required. 
Further, the necessity of such a site, even for the lighter 
$n$-capture elements 
below barium,
has also  been called into question.
It has been proposed 
that both
ends of the abundance distribution could be synthesized in
different regions of the same  neutron-rich jet of a core-collapse
supernova (Cameron 2001)\nocite{cam01}.
Clearly, additional stellar observational data for elements below barium
will be important in helping to resolve this question.

As seen in Figure~\ref{f10}
 the abundances of the elements Sr, Y and Zr in
\bd17\ also do
not follow the solar $r$-process curve that matches the heavier $n$-capture
elements. This is not surprising, as   detailed abundance studies
of several other stars, e.g., 
CS~22892-052 (Sneden et~al. 2000a)\nocite{sne00a}, 
HD 115444 (Westin et~al. 2000)\nocite{wes00} and 
HD 122563 \cite{sne98}, have all shown this disconnect. 
The production of the elements, Sr, Y and Zr, however, may be  complicated 
since they  
could be produced not only in the $r$-process but in the weak $s$-process 
occuring in massive stars \cite{rat93}.
Thus, some combination of those processes could be responsible for 
the abundance pattern we observe in \bd17.  Figure~\ref{f11},
though, suggests that 
more sophisticated $s$-process models, such as those of Arlandini et~al.
(1999\nocite{arl99}), might be more successful in explaining the 
abundances, so far, of Y and Zr in stars such as \bd17. 

We note, finally, that the Ge abundance falls well below the solar curve. 
This element has been seen previously in only three other metal-poor halo
stars, and in all cases the Ge abundance exhibited the same behavior 
(Sneden et~al. 1998)\nocite{sne98}.  Interestingly,  in those other cases the 
Ge abundance seems to scale, at least weakly, with the iron abundance.
Thus, the Ge abundance in both HD 115444 and HD 122563, stars
with [Fe/H] $\simeq$ --3, was the same. HD 126238, with a factor of
about 20 higher in metallicity, has a Ge abundance about a factor of 
16 higher than in the first two stars. The Ge abundance for \bd17, with
[Fe/H] = --2, is  consistent with that pattern,
falling between those of HD 115444 and HD 122563 and the more metal-rich
HD 126238. The scaling is not exact, however.  We are currently 
examining this   
possible weak metallicity dependence,  as well as  
possible nucleosynthetic explanations  for this behavior  
(Truran et~al. 2001, in preparation). 

\subsection {Radioactive Age Determinations}

Our new observations indicate the presence of thorium in \bd17.
This adds to the growing list of such Th detections in metal-poor and 
UMP stars,  e.g., SMPCBA, Sneden et~al. (2000a)\nocite{sne00a},
Westin et~al (2000)\nocite{wes00}, Johnson \& Bolte (2001)\nocite{jb01},
Cayrel et~al (2001)\nocite{cay01}.
These detections, therefore, allow for chronometric stellar age
estimates, typically by comparing 
the observed stellar abundance ratio
Th/Eu  
with predictions of the  initial (zero-decay) value of
Th/Eu at the time of the formation of these elements.
(Eu is employed since it is also a predominantly {\it r}-process element and
has spectral features that are easily observed
in most metal-poor stars.)

The very recent, and first, detection of uranium in the UMP star 
CS~31082-001 ([Fe/H] = --2.9)  by Cayrel et~al. (2001) offers  promise 
for these stellar age detections with the addition of a second chronometer.
The spectrum of  \bd17\ indicates the presence of uranium, albeit weakly,
(see
Figure~\ref{f9}), 
marking this as the second metal-poor halo star in which U has been
detected. 
Such detections may,  however,  turn out to be uncommon as previous observations
have failed to detect U in such stars as CS~22892--052 and HD~115444
(see also Burles et~al. 2001 for further discussion).

Cayrel et~al. (2001) (see also 
Toenjes et~al. 2001) 
employed the abundances of U and Th, in combination 
with each other and with some other stable elements, 
to find an average chronometric age of 12.5  $\pm$ 3 Gyr 
for CS~31082-001.
To make these age calculations they relied mostly upon 
the predicted initial (zero-decay) values listed in Cowan et~al.
(1999), based upon the 
extended Thomas Fermi
model with quenched shell effects far from stability, i.e., 
ETFSI-Q (see Pfeiffer et~al. 1997 for discussion.)
Predictions for initial abundance ratios of Th/Eu, or Th to other stable
elements, have been necessary since 
the nuclei involved in the {\it r}-process are far from stability, 
and nuclear data have not in general been available.
That situation has been slowly changing, however, with more nuclear
data leading to increasingly more reliable 
prescriptions for very
neutron-rich  nuclei
(see Pfeiffer et~al. 2001a; Burles et~al. 2001). 

We have employed the Th/U abundance ratio, similarly to Cayrel et~al.,
to make age calculations for \bd17. 
The uranium detection in \bd17, however, is quite weak and has a 
relatively large error bar. One possible interpretation of our
data is that we are, 
in fact, only  measuring the upper limit on that abundance, 
an alternative we consider below. 
As in other previous studies,
we have also included chronometric age estimates for \bd17 
based upon the Th/Eu
ratio. 
In addition, our HST observations have provided the abundances of 
the important \third\ $r$-process peak elements, particularly Pt. 
These (heaviest stable) 
elements are closer in nuclear mass number to Th and U, than Eu for 
example, and thus offer a direct comparison with ratios 
of Th or U to less massive $n$-capture elements.
Similarly to Cayrel et~al. we have made use of the predicted initial
(zero-decay) abundance ratios from Cowan et~al. (1999), based in all
but one case on the ETFSI-Q (least square) model calculations.
For the case of Th/Eu only, we have followed Cowan et~al. 
and adopted an average value based upon the ETFSI-Q, ETFSI-Q (least
square) and finite range droplet with microscopic
shell corrections  (FRDM+HFB)  models.
Further discussion of the reliability of the various nuclear mass formulae,
along with new theoretical $r$-process calculations and new initial abundance 
ratios are presented in Burles et~al. (2001).

In Table~\ref{tab4} we list the chronometer pair, the 
predicted (initial, zero-decay) abundances, the observed abundances 
and the corresponding
predicted age, based upon a direct comparison between the predicted
and observed ratios. 
We have not used the uncertain Os  abundance to make
these predictions, and we have deliberately  
grouped the uranium chronometers separately to 
emphasize that the weak uranium detection has much larger error
bars associated with it than the thorium abundance does. Nevertheless,
we have increased the number of chronometer pairs by a factor of six 
over most earlier studies, which only employed Th/Eu. 
(Cayrel et~al. 2001 listed  three pairs for their study of 
CS~31082-001, and Hannawald, Pfeiffer, \& Kratz 2001 derived a 
new age estimate for 
CS~22892--052 from weighted averages of 15 Th/Ba-Ir pairs.)
Our results demonstrate 
a general consistency  between the various chronometer-pair
estimates, with a mean value of 13.8~$\pm$~4~Gyr.  
(We have also not employed  the uncertain gold abundance for age
estimates, but comparisons
between predicted and observed Th/Au and U/Au suggest ages of 9.5 and 
12.1 Gyr, respectively.)
As an additional check on the age estimates,  we have also included in 
Table~\ref{tab4} the solar system value for the different chronometer
pairs. These solar $r$-process values are from BPASCR 
and are not dependent upon nuclear mass models for the abundance
ratios.
The last column in the table lists age estimates based upon comparing the
solar ratios with the observed stellar ratios. These ages  should be 
viewed as  
lower limits, 
since the solar values (at 4.5 Gyr old) are lower limits on  the
zero decay-age $r$-process abundances. (Eu is stable but Th 
has partially decayed.) There are some  differences in the 
ages based upon different chronometer pairs,
and between the theoretically predicted and solar values. We note, 
for example, that some of the  
predicted ages are less than
the expected lower limit solar system age estimates. This is probably 
an indication that some of the predicted values from Cowan et~al. (1999) 
need to be revised (see Burles et~al. 2001).
Nevertheless, 
the (lower limit) age 
derived from  the 
solar abundance  ratios, 
13.2 Gyr, is not inconsistent with the average based upon 
theoretical predicted $r$-process abundance ratios.
Since the uranium detection is so weak, we could also consider our
abundance measurement for this element as an upper limit rather than
a detection. In this alternative the age estimates using this 
chronometer would more properly be lower limits.
We have noted that possibility explicitly (by including $\ge$ signs)  in 
Table~\ref{tab4}  
for the age estimates employing uranium.

Our derived age for \bd17\ (13.8~$\pm$~4~Gyr) is in agreement with the 
cosmochronological ages for CS~31082-001 (12.5~$\pm$~3~Gyr, Cayrel et~al. 
2001) and for CS~22892--052 and HD 115444 (15.6~$\pm$~4~Gyr,
Cowan et~al. 1999; Westin et~al. 2000).  
We caution, however,  that all of these age estimates 
are very sensitive to uncertainties both in the theoretically predicted initial
values and in the observations themselves -- this is particularly
true for our very weak detection of uranium. 
In addition, further investigation of any possible real offset
between  the rare-earth elements and the \third\ $r$-process 
peak elements and the corresponding effect on 
nucleocosmochronometry will be necessary.
On the other hand, the overlap in age estimates found so far does 
seem promising and we note that this  technique has the advantage of 
being independent of (uncertain) models of 
Galactic chemical evolution.

\section{CONCLUSIONS}

We have gathered extensive ground-based (Keck and McDonald Observatory)
and space-based (HST) high-resolution spectra of the metal-poor Galactic
halo star \bd17.
Our observations have detected the \third\ $r$-process peak elements,
Os, Ir and Pt, making \bd17\ one of the few stars 
where these elements have been positively identified, and the only
star with detectable gold.
In addition,  we have tentatively detected uranium
in this star, only the second such
star that indicates  the presence of this element. The detections of elements 
from Ge 
(using HST) up to U, span the widest range (so far) 
in proton number of $n$-capture elements in halo stars.

The abundances of the heavy (Z $\ge$ 56) $n$-capture elements 
in \bd17\ are consistent with a scaled solar system $r$-process abundance
distribution. This result confirms earlier detailed abundance studies
of UMP halo stars. Further, it suggests a robust nature for the 
$r$-process throughout the history of the Galaxy. 
However, the abundances of the lighter $n$-capture elements  in the range
40 $<$ Z $<$ 56 (i.e.,  Nb, Pd  and Ag)   
do not fall on the
same  curve as the heavier $n$-capture elements -- 
a result that is consistent with the only
other star with similar abundance information, CS~22892--052. 
This abundance pattern might be produced by two $r$-process sites
(with one for the heavier and one for the lighter
 $n$-capture elements), 
perhaps from
supernovae occurring
at different frequencies,  or neutron star binaries,
or some combination of those.
Alternatively, a single site with two regimes or sets of conditions
in the same supernova site might also offer
an explanation for the abundance pattern seen in this star. 

We have employed the newly detected Th, U and \third\ $r$-process peak 
element abundances to make chronometric age estimates for \bd17. 
The average of the various
chronometric pairs suggests an age of 13.8 $\pm$ 4 Gyr for this star.
This value is consistent, within error limits, with other chronometric 
age determinations for UMP and metal-poor Galactic halo stars. 
The still relatively large uncertainties reflect the sensitivity of 
the age estimate on both the 
predicted and observed abundance ratios. 
It is clear
that additional stellar observations and 
theoretical studies, both already in progress, will be required to 
reduce the uncertainties so that this promising  technique
will be able to provide increasingly accurate age determinations of the 
oldest stars that will put limits on the age of the Galaxy, and 
thus constrain cosmological age estimates.

\acknowledgments

We thank an anonymous referee for helpful comments to improve the paper.
We thank Rica French for assistance with the data reduction and 
Jason Collier for assistance with the solar system abundances. 
This research has been supported in part by STScI grants GO-8111 and GO-08342,
NSF grants  
AST-9986974  (JJC),  
AST-9987162  (CS),   
AST-0098508 and AST-0098549  (TCB), 
AST-9819400 (JEL),  PHY-9800980 (GMF),
and by the ASCI/Alliances Center for Astrophysical Thermonuclear Flashes 
under DOE contract B341495 (JWT). 
Support was also provided by 
the German  BMBF (grant 06MZ9631).

\newpage

\appendix

\section{LINE DATA ADOPTED FOR THE ABUNDANCE ANALYSES}

In this appendix we comment on the selection of atomic transition data for
elements with Z~$>$ 30 and on the details of
the line lists constructed for the synthetic spectrum computations.
We discuss here  only those elements for which CS~22892-052
    abundances were given by Sneden et~al. (2000a) without citation of atomic
    data sources, and those for which different atomic data choices have been
    made than those already discussed in previous papers of this series.
Thus the reader is referred to SMPCBA for a discussion of
\ion{Sr}{1}, \ion{Sr}{2}, \ion{Y}{2}, \ion{Zr}{2}, \ion{Ba}{2}, 
\ion{Sm}{2}, \ion{Ho}{2}, \ion{Er}{2}, and \ion{Tm}{2}, to Sneden et~al. 
(1998) for some additional information on \ion{Os}{1}, \ion{Pt}{1}, and
\ion{Pb}{1}, and to Crawford et~al. (1998)\nocite{CSKBD98} for \ion{Ag}{1}.
We also recomputed partition functions for most of the $n$-capture species
    studied here, instead of relying solely on those computed
    by Irwin (1981)\nocite{Ir81}, which have become out-of-date for
    especially some ionized rare-earth species.

{\em \ion{Ge}{1}, Z = 32:} The 3039.07~\AA\ line is the principal Ge abundance
indicator, and it has been included in several laboratory analyses 
(see Sneden et~al. 1998 for brief discussion and references).
Here we adopt the $gf$ from the recent study of Bi\'emont et~al. (1999),
averaging their experimental and theoretical values.
There are a few other \ion{Ge}{1} lines that should be detectable on
our HST-STIS spectrum of \bd17, and we attempted to synthesize
the ones at 2651.17, 2651.57~\AA\ as part of the modeling effort
for the \ion{Pt}{1} 2650.85~\AA\ line.
Both of the lines proved however to suffer significant contamination
from other atomic and OH features.
    In spite of these problems, it is possible to state that the profiles
    of these lines
will not admit a large Ge
abundance: we estimate that log~$\epsilon$~$\lesssim$ +0.9 from the 
2651.17~\AA\ line and $\lesssim$ +0.6 from the 2650.85~\AA\ line.
These values generally support the abundance derived from the 
more trustworthy 3039~\AA\ line.

{\em \ion{Nb}{2}, Z = 41:} the $gf$ values are taken from Hannaford
    et~al. (1985)\nocite{HLBG85}.

{\em \ion{Pd}{1}, Z = 46:} the $gf$ values are taken from Bi\'emont
    et~al. (1982)\nocite{BGKZ82}.

{\em \ion{La}{2}, Z = 57:} a new laboratory study of this species by Lawler,
Bonvallet, \& Sneden (2001a)\nocite{lbs01} provides transition
probabilities and hyperfine structure constants for a large number of 
lines.  \ion{La}{2}  has a wide hyperfine pattern for many 
transitions, but La has just one naturally occurring isotope 
($^{139}$La), so isotopic wavelength shifts are not needed.
From these data we found 15 useful transitions in the \bd17\ spectrum,
double the number employed by SMPCBA for CS~22892-052.
The old line lists were refined with the new data, and line lists
for the new transitions were created according to the prescriptions
given in the earlier paper.

{\em \ion{Ce}{2}, Z = 58:} we adopted the extensive set of new $gf$ values
of Palmeri et~al. (2000)\nocite{PQWB00}, who combined results from both
experimental and theoretical investigations.
We detected 22 lines of this species (compared with nine employed by 
SMPCBA for the CS~22892-052 analysis).
Since all \ion{Ce}{2} lines are fairly weak in the \bd17\
spectrum, log(EW/$\lambda$)~$\lesssim$ --5.3, we were able to treat them
as single lines without hyperfine or isotopic splitting.

{\em \ion{Pr}{2}, Z = 59:} this species has been reconsidered by Ivarsson
et~al. (2001)\nocite{ivar}, who have generated improved sets of 
energy levels, transition probabilities, and hyperfine structure 
constants for many lines.
Since their data should be internally consistent,   we limited
our Pr abundance analysis to just those lines considered by Ivarsson
et~al., in preference to the atomic data chosen by SMPCBA.
As is the case for La, Pr has only one stable isotope ($^{141}$Pr).
After identifying \ion{Pr}{2} lines in the \bd17\ spectrum we generated
entirely new linelists for synthetic spectrum computations.

{\em \ion{Nd}{2}, Z = 60:} this species lacks a recent laboratory study.
    SMPCBA attempted to reconcile the $gf$ scales of Ward et~al.
    (1985)\nocite{WVAHW85} and Maier \& Whaling (1977)\nocite{MW77},
    but the line-to-line scatter in $gf$ values between these two studies
    is large (Sneden, Lawler, \& Cowan 2001)\nocite{SLC01}.
    Moreover, these laboratory studies only consider a small subset of
    the \ion{Nd}{2} lines that can easily be observed in stellar spectra.
    Therefore to produce $gf$ values for our lines we simply averaged the
    large-sample, medium-accuracy $gf$ results of 
Cowley \& Corliss (1983) 
with those of Ward et~al. 
    and Maier \& Whaling
    where available.
    This procedure allowed us to measure a large set of \ion{Nd}{2} lines,
    but the resulting Nd abundance has a larger internal uncertainty
    ($\sigma$~= 0.17) than most other rare-earth elements.

{\em \ion{Eu}{2}, Z = 63:} 
new experimental $gf$ values and hyperfine/isotopic splitting
    constants have been determined by Lawler et~al. 
(2001c)\nocite{lwds01}, who also summarize the line data for this species
from other relevant recent studies.
Complete line parameters are now available for all strong \ion{Eu}{2}
transitions.

{\em \ion{Gd}{2}, Z = 64:} as in the SMPCBA study, we employed
the $gf$'s of Bergstr{\"o}m et~al. (1988)\nocite{BBLP88} that
were derived directly from their lifetime measurements.
For other lines, we adopted the values of Corliss \& Bozman 
(1962)\nocite{CB62}, multiplied by 1.3 as recommended by Bergstr{\"o}m et~al.
The proposed re-normalization of the older $gf$ scale really only
applies to transitions connecting to those levels for which 
Bergstr{\"o}m et~al. measured lifetimes.
However, derivation of Gd abundances using only the six measured lines 
in \bd17\ in common with the Bergstr{\"o}m et~al. study yielded nearly
the same mean abundance as computed from the whole set of 15 lines,
and the sample standard deviations $\sigma$ were identical.
Therefore all \ion{Gd}{2} lines were given equal weight in the abundance
analysis.

{\em \ion{Tb}{2}, Z = 65:} poorly determined transition parameters and 
intrinsic weakness of all lines of this species in the Sun and metal-poor 
stars in the past have led to very uncertain abundances of Tb.
New laboratory $gf$ and hyperfine structure data ($^{159}$Tb is the only
stable isotope of this element, so isotopic wavelength shifts are
not needed) have been obtained by Lawler et~al. (2001b)\nocite{lwcs01} 
for \ion{Tb}{2}.
These data, along with updated partition functions, have been used
to analyze the five \ion{Tb}{2} lines that could be 
cleanly detected in the \bd17\ spectrum.  These lines yield
    very consistent Tb abundances in this star.

{\em \ion{Dy}{2}, Z = 66:} there are several laboratory analyses of this 
species, and for maximum internal consistency we adopted only the $gf$ values 
from the recent study of Wickliffe, Lawler \& Nave (2000)\nocite{WLN00}.
However, use of the Kusz (1992)\nocite{Kus92} or Bi\'emont \& Lowe 
(1993)\nocite{BL93} $gf$'s would not materially alter the Dy abundance
derived for \bd17, since all three lab studies find similar results 
for the stronger lines of \ion{Dy}{2}.

{\em \ion{Os}{1}, Z = 76:} the $gf$ values have been taken from 
Kwiatkowski et~al. (1984)\nocite{KZBG84}, or scaled from those of
Corliss \& Bozman (1962)\nocite{CB62} according to the lifetime
measurements of Kwiatkowski et~al.
The resulting abundances derived from five lines (two from Keck-HIRES
and three from HST-STIS spectra) are in excellent internal agreement.
However, all \ion{Os}{1} lines are weak in \bd17. 
Moreover, the 2838.63~\AA\ and 3301.57~\AA\ features are particularly 
blended with other atomic lines, and the S/N of the Keck-HIRES spectrum
in the region of the 3267.95~\AA\ line is not high; the reader should
exercise caution in interpretation of the Os abundance.

{\em \ion{Ir}{1}, Z = 77:} The $gf$'s are based on the lifetime
measurements of Gough, Hannaford, \& Lowe (1983)\nocite{GHL83}.

{\em \ion{Pt}{1}, Z = 78:} using HST-GHRS spectra of a few very small 
wavelength intervals, Sneden et~al. (1998)\nocite{sne98} synthesized the 
\ion{Pt}{1} 2929.79, 3064.71~\AA\ lines to derive abundances of Pt in the 
$r$-process-rich star HD~115444 and in the higher metallicity giant
HD~126538 ([Fe/H]~= --1.65, [$n$-capture/Fe]~$\sim$~0).
The 3064~\AA\ line is not easy to work with as it is nearly buried in
a thicket of other atomic and OH molecular contaminants.
The much larger wavelength range of the STIS spectrum allowed detection
of five \ion{Pt}{1} lines in \bd17, including the very strong one at
2659.45~\AA.
We employed the $gf$ values recommended by Morton (2000)\nocite{Mor00},
based on experimental results of Gough, Hannaford, \& Lowe 
(1982)\nocite{GHL82}, Lotrian \& Guern (1982)\nocite{LG82}, and
Reader et~al. (1990).

{\em \ion{Au}{1}, Z = 79:} the derived Au abundance should be
    interpreted with care.
    This abundance is determined from the two resonance lines at
    2427.94 and 2675.94~\AA, whose $gf$ values are well-determined
    (Hanneford, Larkin, \& Lowe 1981\nocite{HLL81}, Gaarde et~al.
    1994\nocite{GZCZLS94}).
    These lines were clearly detected in the spectrum of \bd17, but
    both suffer substantial blending problems.
    Discussion of the 2675~\AA\ line was given in \S3.3.
    Derivation of an Au abundance from the 2427.94~\AA\ line is more
    problematic, because it sits on the side of a very strong \ion{Fe}{1}
    line at 2428.06~\AA, and only a very rough guesstimate of
    log~$\epsilon$(Au)~$\sim$ --0.2 could be made, with an uncertainty
    of at least $\pm$0.5~dex.

{\em \ion{U}{2}, Z = 92:} the $gf$ of the 3859.57~\AA\ was taken from 
the new study by Lundberg et~al. (2001)\nocite{LJNZ01}.

\begin{table}
\dummytable\label{tab1}
\end{table}

\begin{table}
\dummytable\label{tab2}
\end{table}

\begin{table}
\dummytable\label{tab3}
\end{table}

\begin{table}
\dummytable\label{tab4}
\end{table}

\newpage 

\newpage 

\newpage

\newpage


\clearpage


\begin{thebibliography}{}


\bibitem[Alonso et al. 1998]{aam98}
Alonso, A., Arribas, S., \& Martinez-Roger, C. 1998, \aaps, 131, 209

\bibitem[Alonso et al. 1998]{aam99}
Alonso, A., Arribas, S., \& Martinez-Roger, C. 1999, \aaps, 140, 261

\bibitem[Alonso et al. 2001]{aam01}
Alonso, A., Arribas, S., \& Martinez-Roger, C. 2001, A\&A, 376, 1039 

\bibitem[Anthony-Twarog \& Twarog 1994]{att94}
Anthony-Twarog, B. J., \& Twarog, B. A. 1994, \aj, 107, 1577

\bibitem[Anthony-Twarog \& Twarog 1994]{att98}
Anthony-Twarog, B. J., \& Twarog, B. A. 1998, \aj, 116, 1922

\bibitem[Aoki et al. 2001]{aok01}
Aoki, W., Ryan, S.G., Norris, J.E., Beers, T.C., Ando, H., Iwamoto, N., Kajino,
T., Mathews, G.J., \& Fujimoto, M.Y. 2001, ApJ, 561, 346.

\bibitem[Arlandini et al. 1999]{arl99}
Arlandini, C., K\"appeler, F., Wisshak, K., Gallino, R., Lugaro, M., Busso, M.,
\& Staniero, O. 1999, \apj, 525, 886

 \bibitem[Balachandran \& Carney 1996]{BC96}
    Balachandran, S., \& Carney, B. W. 1996, \aj, 111, 946

\bibitem[Barlow \& Sargent]{bar97}
Barlow, T. A. \& Sargent, W. L. W. 1997, AJ, 113, 136

\bibitem[Beers et al. 1985]{ber85}
Beers, T. C., Preston, G. W., \& Shectman, S. A. 1985, AJ, 90, 2089

\bibitem[Beers et al. 1992]{ber92}
Beers, T. C., Preston, G. W., \& Shectman, S. A. 1992, AJ, 103 1987

\bibitem[Beers et al. 1999]{brnrs99}
Beers, T. C., Rossi, S., Norris, J. E., Ryan, S. G., \& Shefler, T. 
1999, \aj, 117, 981

\bibitem[Beers \& Sommer-Larsen 1995]{BS95}
Beers, T. C., \& Sommer-Larsen, J. 1995, \apjs, 96, 175

\bibitem[Bell \& Gustafsson 1978]{bg78}
Bell, R. A., \& Gustafsson, B. 1978, \aaps, 34, 229

\bibitem[Bergstr{\"o}m {\it et~al.} 1988]{BBLP88}
Bergstr{\"o}m, H., B\'iemont, E., Lundberg, H., \& Persson, A.  1988,
\aap, 192, 335

\bibitem[Bi\'emont et al. 1982]{BGKZ82}
    Bi\'emont, E., Grevesse, N., Kwiatkowski, M., \& Zimmermann, P.
    1982, \aap, 108, 127

\bibitem[Bi\'emont & Lowe 1993]{BL93} 
Bi\'emont, E. \& Lowe, R. M. 1993, \aap, 273, 665

\bibitem[Bi\'emont et al. 1999]{BLLSGD99}
Bi\'emont, E., Lyng{\aa}, C., Li, Z. S., Svanberg, S., Garnir, H. P., 
\& Doidge, P. S. 1999, \mnras, 303, 721

   \bibitem[Boesgaard et al. 1999]{BKDV99}
    Boesgaard, A. M., King, J. R., Deliyannis, C. P., \& Vogt, S. S.
    1999, \aj, 117, 492

\bibitem[Bond 1980]{bon80}
Bond, H. E. 1980, Ap.J.Suppl., 44, 517

\bibitem[Burles et al. 2001]{bur01}
Burles, S., Truran, J. W., Cowan, J. J., Sneden, C., Kratz, K.-L, 
\& Pfeiffer, B. 2001, preprint

\bibitem[Burris et al. 2000]{bur00}
Burris, D. L., Pilachowski, C. A., Armandroff, T. A., Sneden, C., Cowan, J. J.,
\& Roe, H. 2000, \apj, 544, 302 (BPASCR) 

\bibitem[Cameron 2001]{cam01}
Cameron, A.G.W. 2001, \apj, in press 

\bibitem[Carbon et al. 1982]{Cetal82}
Carbon, D., Langer, G., Butler, D., Kraft, R., Suntzeff, N., Kemper, E.,
Trefzger, C., and Romanishin, W.  1982, \apjs, 49, 207

\bibitem[Carney 1983]{car83}
Carney, B. W. 1983, \aj, 88, 610

\bibitem[Cayrel 1996]{cay96}
Cayrel, R. 1996, \aapr, 7, 217

\bibitem[Cayrel et al. 2001]{cay01} 
Cayrel, R., et al. 2001,  Nature,  409, 691


\bibitem[Clari\'a et al. 1994]{cmpl94}
Clari\'a, J. J., Minniti, D., Piatti, A. E., \& Lapasset, E. 1994, \mnras, 
268, 733

\bibitem[Corliss & Bozman 1962]{CB62}
Corliss, C. H. \& Bozman, W. R. 1962, NBS Monograph 32 (Washington: 
US Gov. Prt. Off.)

\bibitem[Cowan et al. 1995]{cow95}
Cowan, J. J., Burris, D. L., Sneden, C., McWilliam, A., \& Preston, G. W.
1995, \apjl, 439, L51

\bibitem[Cowan et al. 1997]{cow97}
Cowan, J. J.,  McWilliam, A.,  Sneden, C., \&  Burris, D. L.
1997, \apj,  480, 246 

\bibitem[Cowan et al. 1999]{cow99}
Cowan, J. J., Pfeiffer, B., 
Kratz, K.-L., Thielemann, F.-K., Sneden, C., Burles, S.,
Tytler, D., \& Beers, T. C. 1999, \apj, 521, 194 

\bibitem[Cowan et al. 1996]{cow96} Cowan, J. J., Sneden, C., Truran,
J. W., \& Burris, D. L. 1996, \apjl, 460, L115

\bibitem[Cowan et al. 2001]{cow01}
Cowan, J. J.,  Sneden, C., \& Truran, J. W. 2001,
in 
{\it Proceedings of the 20th Texas Symposium on
Relativistic Astrophysics}, ed. C. Wheeler \& H. Martel,  astro-ph/0103222,
in press  

\bibitem[Cowley \& Corliss 1983]{CC83}
Cowley, C. R., \& Corliss, C. H. 1983, \mnras, 203, 651

\bibitem[Crawford et al. 1998]{CSKBD98} 
Crawford, J. L., Sneden, C., King, J. R., Boesgaard, A. M., \& Deliyannis,
C. P. 1998, \aj, 116, 2489

\bibitem[Dupree \& Smith 1988]{ds88}
Dupree, A. K., \& Smith, G. H. 1988, \aj, 95, 1547

\bibitem[Dupree \& Smith 1995]{ds95}
Dupree, A. K., \& Smith, G. H. 1995, \aj, 110, 405


\bibitem[Edvardsson et al. 1993]{edv93}
Edvardsson, B., Andersen, J., Gustafsson, B., Lambert, D. L., Nissen, P. E.,
\& Tomkin.J. 1993, A\&A, 275, 101

\bibitem[Fitzpatrick & Sneden 1987]{FS87}
Fitzpatrick, M.~J. \& Sneden, C.  1987, \baas, 19, 1129

\bibitem[]{FSS93}
Fran\c{c}ois, P., Spite, M., \& Spite, F. 1993, \aap, 274, 821

\bibitem[Gaarde et al. 1994]{GZCZLS94}
Gaarde, M. B., Zerne, R., Caiyan, L., Zhankui, J., Larsson, J.,
\& Svanberg, S. 1994, Phys. Rev. A., 50, 209

\bibitem[Gilliland et al. 1992]{GMWEL92}
Gilliland, R. L., Morris, S. L., Weymann, R. J., Ebbets, D. C., 
\& Linder, D. J. 1992, \pasp, 104, 367

\bibitem[Gilroy et al. 1988]{gil88}
Gilroy, K. K., Sneden, C., Pilachowski, C. A.,  \&  Cowan, J. J. 1988,
\apj, 327, 298

\bibitem[Gough et al. 1982]{GHL82}
Gough, D. S., Hannaford, P., \& Lowe, R. M. 1982, J. Phys. B., 15, L431

\bibitem[Gough et al. 1982]{GHL83}
Gough, D. S., Hannaford, P., \& Lowe, R. M. 1982, J. Phys. B., 16, 785

\bibitem[Gratton \& Sneden 1991]{gra91}
Gratton, R., \& Sneden, C. 1991, \aap, 241, 501

\bibitem[Gratton \& Sneden 1994]{gra94}
Gratton, R., \& Sneden, C. 1994, \aap, 287, 927

\bibitem[Gratton et al. 2000]{gscb00}
Gratton, R. G., Sneden, C., Carretta, E., \& Bragaglia, A. 2000, 
\aap, 354, 169

\bibitem[Grevesse \& Sauval 1998]{GS98}
Grevesse, N., \& Sauval, A. J., 1998, Sp. Sci. Rev., 85, 161

\bibitem[Hannaford et al. 1981]{HLL81}
Hannaford, P., Larkins, P. L., \& Lowe, R. M. 1981, J. Phys. B., 14, 2321

\bibitem[Hannaford et al. 1985]{HLBG85}
    Hannaford, P., Lowe, R. M., Bi\'emont, E., \& Grevesse, N. 1985,
    \aap, 143, 447

\bibitem[Hannawald et al. 2001]{han01}
Hannawald, M., Pfeiffer, B., \& Kratz, K.-L. 2001, 
in  Astrophysical Ages and Time Scales,
ASP Conference Series, eds. T. von Hippel, N. Manset \& C. Simpson, 310

\bibitem[]{HY82}
Hartkopf, W. I., \& Yoss, K. M. 1982, \aj, 87, 1679

\bibitem[Holweger & M{\"u}ller 1974]{HM74}
Holweger, H. \& M{\"u}ller, E. A.  1974, \solphys, 39, 19.

\bibitem[Irwin 1981]{Ir81}
Irwin, A. W. 1981, \apjs, 45, 621

\bibitem[Israelian et al. 1998]{IGR98}
   Israelian, G., Garc\'ia Lopez, R. J., \& Rebolo, R. 1998, \apj, 507, 805

\bibitem[Ivarsson et al. 2001]{ivar}
Ivarsson, S., Litz\'en, U., \& Wahlgren, G. M., 2001, Phys. Scr., in press

\bibitem[Johnson \& Bolte 2001]{jb01}
Johnson, J. A., \& Bolte, M. 2001, \apj, 554, 888

\bibitem[K\"appeler et al. 1989]{kap89}
K\"appeler, F., Beer, H., \& Wisshak, K. 1989, Rep. Prog. Phys.,
52, 945

\bibitem[Klochkova et al. 1999]{kep99}
Klochkova, V. G., Ermakov, S. V., \& Panchuk, V. E. 1999, \apss, 265, 185

\bibitem[Kraft et al. 1982]{ketal82}
Kraft, R. P., Suntzeff, N. B., Langer, G. E., Trefzger, C. F., Friel, E.,
Stone, R. P., \& Carbon, D. F. 1982, \pasp, 94, 55

\bibitem[Kurucz 1995]{Ku95}
Kurucz, R. L. 1995, in Workshop on Laboratory and astronomical high 
resolution spectra, ASP Conference Ser. \#81 ed. A.J. Sauval, R. Blomme, 
and N. Grevesse (San Francisco: Astr. Soc. Pac.), p.583

\bibitem[Kurucz {\it et~al.} 1984]{KFBT84}
Kurucz, R.~L., Furenlid, I., Brault, J., \& Testerman, L.  1984,
Solar Flux Atlas from 296 to 1300 nm (Cambridge, MA: Harvard Univ.).

\bibitem[Kusz 1985]{Kus92} 
Kusz, J. 1992, \aaps, 92, 517

\bibitem[Kwiatkowski et al. 1984]{KZBG84}
Kwiatkowski, M., Zimmermann, P., Bi\'emont, E., \& Grevesse, N.
1984, \aap, 135, 59

\bibitem[Lawler et al. 2001a]{lbs01}
Lawler, J. E., Bonvallet, G., \& Sneden, C. 2001a, \apj, 556, 452

\bibitem[Lawler et al. 2001b]{lwcs01}
Lawler, J. E., Wickliffe, M. E., Cowley, C. R., \& Sneden, C. 2001b,
ApJS, in press

\bibitem[Lawler et al. 2001c]{lwds01}
Lawler, Wickliffe, M. E., Den Hartog, E. A., \& Sneden, C. 2001c, \apj,
in press

\bibitem[Lotrian \& Guern 1982]{LG82}
Lotrian, J., \& Guern, Y. 1982, J. Phys. B., 15, 69

\bibitem[Luck \& Bond 1981]{LB81}
Luck, R. E., \& Bond, H. E. 1981, \apj, 244, 919

\bibitem[Lundberg et al. 2001]{LJNZ01}
Lundberg, H., Johansson, S., Nilsson, H., \& Zhang, Z. 2001, \aap, 372, 50

\bibitem[Maier & Whaling 1977]{MW77}
Maier, R. S. \& Whaling, W. 1977, J. Quant. Spec. Rad. Trans., 18, 501

\bibitem[Mathews, Bazan, \& Cowan 1992]{mat92}
Mathews, G. J., Bazan, G., \& Cowan, J. J. 1992,  \apj, 391, 719

\bibitem[McWilliam 1998]{mcw98}
McWilliam, A. 1998, AJ, 115, 1640

\bibitem[McWilliam {\it et~al.} 1995a]{MPSS95a}
McWilliam, A., Preston, G. W., Sneden, C., \& Shectman, S.  1995a,
\aj, 109, 2736.

\bibitem[McWilliam {\it et~al.} 1995b]{MPSS95b}
McWilliam, A., Preston, G.~W., Sneden, C., \& Searle, L.  1995b,
\aj, 109, 2757.

\bibitem[Mel\'endez et al. 2001]{MBS01}
Mel\'endez, J., Barbuy, B., \& Spite, F. 2001, New Astr. Rev., 45, 551

\bibitem[Moore {\it et~al.} 1966]{MMH66}
Moore, C.~E., Minnaert, M. G. J., \& Houtgast, J.  1966,
The Solar Spectrum 2935{\AA} to 8770{\AA}, NBS Mono. 61

\bibitem[Morell et al. 1992]{MKB92}
Morell, O., K{\"a}llander, D., \& Butcher, H. R. 1992, \aap, 259, 543

\bibitem[Morton 2000]{Mor00}
Morton, D. C. 2000, \apjs, 130, 403

\bibitem[]{NRB97}
Norris, J. E., Ryan, S. G., \& Beers, T. C., 1997, \apj, 489, L169

\bibitem[Palmeri et al. 2000]{PQWB00}
Palmeri, P., Quinet, P., Wyart, J.-F., \& Bi\'emont, E., 2000, Phys. Scr., 
61, 323

\bibitem[Perryman et al. 1997]{per97}
Perryman, M. A. C. et al. 1997, \aap, 323, L49

\bibitem[Pfeiffer et al. 1997]{pfe97}
Pfeiffer, B., Kratz, K.-L., \& Thielemann, F.-K. 1997, Z. Phys. A, 357, 235

\bibitem[Pfeiffer et al. 2001a]{pfe01a}
Pfeiffer, B.,  Kratz, K.-L.,  Thielemann, F.-K., \&   Walters, W. B. 
2001a, Nucl. Phys. A693,  282 

\bibitem[Pfeiffer et al. 2001b]{pfe01b}
Pfeiffer, B., Ott, U., \&  Kratz, K.-L.
2001b, Nucl. Phys. A688,  575 

\bibitem[Pilachowski et al. 1996]{psb93}
Pilachowski, C. A., Sneden, C., \& Booth, J. 1993, \apj, 407, 699

\bibitem[Pilachowski \& Sneden]{pse99}
Pilachowski, C. A., \& Sneden, C.  1999, AAS, 195, 5006

\bibitem[Pilachowski et al. 1996]{psk96}
Pilachowski, C. A., Sneden, C., \& Kraft, R. P. 1996, AJ, 111, 1689

\bibitem[Raiteri et al. 1993]{rat93}
Raiteri, C. M., Gallino, R., Busso, M., Neuberger, D. \& K\"appeler, F.
 1993, \apj, 419, 207

\bibitem[Reader et al. 1990]{RASS90}
Reader, J., Acquistra, N., Sansonetti, C. J., \& Sansonetti, J. E. 1990
\apjs, 72, 831

\bibitem[Rosswog et al. 1999]{ros99}
Rosswog, S., Liebendorfer, M., Thielemann, F.-K., Davies, M. B., Benz, W.,
\& Piran, T., 1999, \aap, 341, 499

\bibitem[Ryan et al. 1996]{ryan}
Ryan, S. G., Norris, J. E., \& Beers, T. C. 1996, \apj, 471, 254

\bibitem[Sneden 1973]{sne73}
Sneden, C. 1973,  \apj, 184, 839

\bibitem[Sneden et al. 1991]{skpl91}
Sneden, C., Kraft, R. P., Prosser, C. F., \& Langer, G. E., 1991,
\aj, 102, 2001

\bibitem[Sneden et al. 1997]{setal97}
Sneden, C., Kraft, R. P., Shetrone, M. D., Smith, G. H., Langer, G. E.,
\& Prosser, C. F. 1997, \aj, 114, 1964

\bibitem[Sneden et al. 2001]{SLC01}
Sneden, C., Lawler, J. E., \& Cowan, J. J. 2001, Phys. Scr., in press

\bibitem[Sneden \& Parthasarathy 1983]{sne83}
Sneden, C., \& Parthasarathy, M. 1983,  \apj, 267, 757

\bibitem[Sneden \& Pilachowski 1985]{sne85}
Sneden, C., \& Pilachowski, C. A. 1985,  \apj,  288, L55

\bibitem[Sneden et al. 2000]{spk00}
Sneden, C., Pilachowski, C. A., \& Kraft, R. P. 2000,  \aj,  120, 1351

\bibitem[Sneden et al. 1991]{sgc91}
Sneden, C., Gratton, R. G., \& Crocker, D. A. 1991, \aap, 246, 354

\bibitem[Sneden et al. 1994]{sne94}
Sneden, C., Preston, G. W., McWilliam, A., \& Searle, L. 1994,
\apj, 431, L27

\bibitem[Sneden et al. 1996]{sne96} 
Sneden, C.,  McWilliam, A., Preston, G. W., Cowan, J. J., Burris, D. L.,
 \& Armosky, B. J. 1996, \apj, 467, 819 (SMPCBA)  

\bibitem[Sneden et al. 2000a]{sne00a} 
Sneden, C., Cowan, J.J., Ivans, I. I., Fuller, G. M., Burles, S., 
Beers, T. C., \& Lawler, J. E.  2000a, \apj, 533, L139.

\bibitem[Sneden et al. 2000b]{sne00b}  
Sneden, C., Johnson, J., Kraft, R. P., Smith, G. H., Cowan, J. J., \& 
Bolte, M. S. 2000b, \apj, 536, L85

\bibitem[Sneden et al. 2001]{sne01a}
Sneden, C.,  Cowan, J. J., \&   Truran, J. W.  2001,
in {\it Cosmic Evolution}, ed. E. Vangioni-Flam and M. Cass\'e,
(Singapore: World Scientific Publishing), astro-ph/0101439, in press

\bibitem[Sneden et al. 2001b]{sne01b}
Sneden, C., Cowan, J. J., 
Lawler, J. E., Burles, S., Beers, T. C., \& Fuller, G. M. 
2001, ApJ, submitted

\bibitem[Sneden et al. 1998]{sne98}
Sneden, C.,  Cowan, J. J., Burris, D. L., \& Truran, J. W.  1998, 
\apj, 496, 235

\bibitem[Spite \& Spite 1978]{spi78}
Spite, M., \& Spite, F. 1978, \aap,   67, 23

\bibitem[Toenjes et al. 2001]{toe01}
Toenjes, R., 
Schatz, H., Kratz, K.-L., Pfeiffer, B.,  Beers, T. C.,
Cayrel, R.,   Cowan, J. J.,  \&   Hill, V.
2001, 
astro-ph/0104335, in  Astrophysical Ages and Time Scales,
ASP Conference Series, eds. T. von Hippel, N. Manset \& C. Simpson, 376 

\bibitem[Truran 1981]{tru81}
Truran, J. W. 1981,  \aap,   97, 391

\bibitem{tru00} Truran, J.W., Fields, B.D., \& Cowan, J.J. 2001, \apj, 
submitted 

\bibitem[Tull et al. 1995]{TMSL95} 
Tull, R. G., MacQueen, P. J., Sneden, C., \& Lambert, D. L. 1995, 
\pasp, 107, 251

\bibitem{Vetal94} Vogt, S. S. et al. 1994, in Proc. SPIE Conf. 2198,
Instrumentation in Astronomy VII, eds. D. L.  Crawford and E. R. Craine, 
(Bellingham, WA: SPIE), p. 362.

\bibitem[Ward {\it et~al.} 1985]{WVAHW85}
Ward, L., Vogel, O., Arnesan, A., Hallin, R., \& W{\"a}nnstr{\"o}m, A. 1985,
Phys. Scr., 31, 161

\bibitem[Wasserburg et al. 1996]{was96}
Wasserburg, G. J., Busso, M., \& Gallino, R. 1996, \apj, 466, L109

\bibitem{was00} 
Wasserburg, G.J. \& Qian, Y.-Z. 2000, \apj, 529, L21

\bibitem[Westin et al. 2000]{wes00} 
Westin, J., Sneden, C., Gustafsson, B., \& Cowan, J.J. 2000, \apj,  530, 783

\bibitem[Wheeler et al. 1998]{whe98}
Wheeler, C., Cowan, J. J., \& Hillebrandt, W.
1998, \apj, 493, L101

\bibitem[Wheeler et al. 1989]{whe89}
Wheeler, C., Sneden, C., \& Truran, J. W.
1989, ARAA, 27, 279 

\bibitem[Wickliffe et al. 2000]{WLN00}
Wickliffe, M. E., Lawler, J. E., \& Nave, G. 2000, \jqsrt, 66, 363

\bibitem[Wisshak et al 1996]{wis96}
Wisshak, K., Voss, F., \& K\"appeler, F. 1996,
in Proceedings of the 8$^{th}$ Workshop on Nuclear Astrophysics,
ed. W. Hillebrandt, \& E. M\"uller ( Munich: MPI), 16

\bibitem[Woolf et al. 199x]{wol95}
 Woolf, V. M., Tomkin, J., \& Lambert D. L. 1995, \apj, 453, 660

\end{thebibliography}
\end{document}